\documentclass[
  journal=jctcce,
  manuscript=article,
  layout=traditional
]{achemso}

\usepackage{graphicx}
\usepackage{amsmath,amssymb}
\usepackage{xcolor}
\usepackage[version=4]{mhchem}
\usepackage{physics}
\usepackage{siunitx}
\AtBeginDocument{\RenewCommandCopy\qty\SI}

\newcommand{\mat}[1]{\ensuremath{\mathbf{#1}}}
\renewcommand{\vec}[1]{\ensuremath{\mathbf{#1}}}

\newcommand{\fig}[1]{Fig.~\ref{#1}}
\newcommand{\refeq}[1]{Eq.~\ref{#1}}

\newif\ifhighlightchanges
\highlightchangesfalse

\newcommand{\princeton}{
  Department of Chemistry,
  Princeton University,
  Washington Road and Scholar Way,
  Princeton, New Jersey 08544, United States
}

\newcommand{\upenn}{
  Department of Chemistry,
  University of Pennsylvania,
  231 S. 34 Street,
  Philadelphia, Pennsylvania 19104, United States
}

\author{D. Vale Cofer-Shabica}
\email{vale@princeton.edu}
\affiliation{\princeton}
\altaffiliation{\upenn} 

\author{Jennifer R. DeRosa}
\affiliation{\princeton}
\altaffiliation{\upenn} 

\author{Joseph E. Subotnik}
\affiliation{\princeton}
\altaffiliation{\upenn} 

\title[]{Marcus Theory and The Condon Approximation Revisited I: E-SHAKE and Seam Sampling}

\begin{document}


\begin{abstract}
Marcus theory is the workhorse of theoretical chemistry for predicting the rates of charge and energy transfer.
Marcus theory overwhelmingly agrees with experiment -- both in terms of electron transfer and triplet energy transfer -- for the famous set of naphthalene-bridge-biphenyl and naphthalene-bridge-benzophenone systems studied by Piotrowiak, Miller, and Closs.  That being said, the agreement is not perfect, and in this manuscript, we revisit one key point of disagreement: the molecule C-13-ae ([3,equatorial]-naphthalene-cyclohexane-[1,axial]-benzophenone).   
To better understand the theory-experiment disagreement, we introduce and employ a novel scheme to sample the seam between two diabatic electronic states (E-SHAKE) through which we reveal the breakdown of the Condon approximation and the presence of a conical  intersection for the C-13-ae molecule; 
we also predict an isotopic effect on the rate of triplet-triplet energy transfer. 
\end{abstract}

\maketitle

\section{Introduction}\label{sec:introduction}
In the nonadiabatic, activated crossing limit (high temperature), the Marcus expression for the rate of electron transfer (ET) is\cite{nitzanCDCP}
\begin{equation} \label{eq:marcus-eet}
  k_{\textrm{D}\to\textrm{A}} =
  \frac{2\pi}{\hbar}
  \abs{H_{\textrm{DA}}}^2
  {\left(4\pi \lambda k_{\textrm{B}}T\right)}^{-1/2}
    e^{-{\left(\lambda + \Delta G^{0}\right)}^2/(4\lambda k_{\textrm{B}}T)}
\end{equation}
Here `D' and `A' indicate donor and acceptor states, and $H_{\textrm{DA}}$ is the electronic matrix element coupling them.
$T$ is the temperature and $k_{\textrm{B}}$ the Boltzmann's constant.
$\Delta G ^{0}$ is the Gibbs free energy associated with the transformation \ce{D \to A} and $\lambda$ describes the contribution of solvent reorganization.
Deriving \refeq{eq:marcus-eet} requires a number of assumptions:
(1) A thermalized, harmonic bath in the
(2) activated crossing limit and
(3) the applicability of golden rule rates \emph{i.e.} the applicability of first order time-dependent perturbation theory.
(4) The Condon approximation further simplifies Eq. \ref{eq:marcus-eet}, treating the electronic coupling as a constant. 
In reality, the coupling depends parametrically on the nuclear coordinate, 
thus the Condon approximation is not necessarily valid for ``floppy'' molecules 
where the crossing region spans vast portions of the nuclear coordinate space (often with wildly different electronic couplings throughout), 
so the utility of the Condon approximation is largely limited to rigid systems.

Ever since the development of Eq. \ref{eq:marcus-eet}, theorists have attempted to go beyond the Condon approximation.
Most importantly, Stuchebrukhov has formulated  a model for electron transfer where one (slow) degree of freedom modulates the energy gap between diabatic states, while another (fast) degree of freedom
(i.e. internal mode) modulates the diabatic coupling; with such a separation, a clean rate expression can be computed.\cite{medvedev1997inelastic}.
Alternatively, non-Condon Hamiltonians where diabatic couplings depend on internal rotational angles have been developed by Jang and Newton.\cite{jang2005theory}.
In general, however, developing realistic models beyond the Condon approximation is difficult both because (i) solving a quantum dynamics problems is usually not tractable and (ii) one never knows how to pick a meaningful form for the geometry dependence of $H_{DA}$.  
Within such a scenario, many theorists today simply choose to run nonadiabatic dynamics \emph{via} fewest-switches surface hopping (FSSH),\cite{tully1990molecular} \emph{ab initio} multiple spawning,\cite{ben2000ab} or other methods. Unfortunately, even for modestly large systems, treating the entire system quantum mechanically becomes untenable---despite the seismic shift in computational capacity with GPU-based computing in the last 10-15 years.\cite{luehr2011dynamic,snyder2016gpu,curchod2017ab,lee2018gpu,seritan2020terachem,kohnke2020gpu,jones2022accelerators,wang2024extending}
Indeed, it is \emph{still} difficult to overstate the predictive utility of such a simple analytic expression like Eq. \ref{eq:marcus-eet}.
For instance, as is well known, the most famous prediction of Eq. \ref{eq:marcus-eet} is the existence of an inverted regime, which was first detected by Closs {\em et. al.} \cite{closs:1986:jcp:inverted}.

\begin{scheme}
    \centering
    \includegraphics[width=0.5\linewidth]{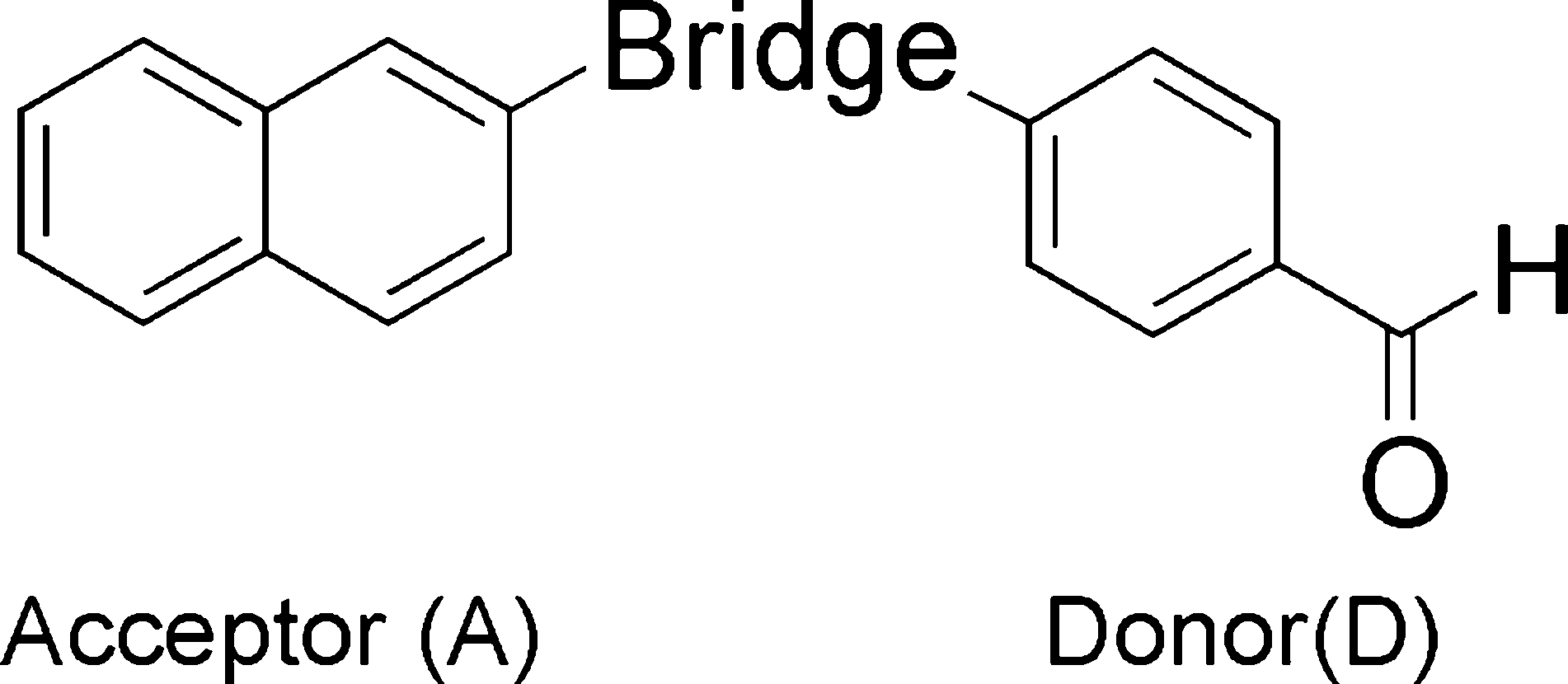}
    \caption{\label{scheme} Donor-bridge-acceptor scheme studied}
\end{scheme}

\begin{figure}
    \centering
    \includegraphics[width=0.5\linewidth]{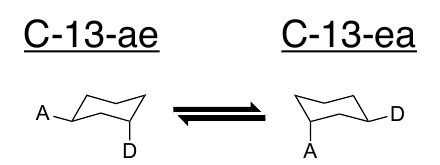}
    \caption{Donor (D) and acceptor (A) locations for C-13-ae and C-13-ea Closs molecules.}
    \label{fig:c13ae_c13ea_interconv}
\end{figure}

Interestingly, while the most famous accomplishment of Ref. \citenum{closs:1986:jcp:inverted} was the confirmation of the existence of the inverted regime for electron transfer, it is worth noting that, over the last 40 years or so, the ``Closs molecules'' continue to be used as a bellwether for testing new theories for electron and \emph{energy} transfer (see Scheme \ref{scheme}). In particular, note that 
while Eq. \ref{eq:marcus-eet} was derived with electron transfer in mind, the formula is a high-temperature expression valid for any two-level nonadiabatic transition.  Indeed, shortly after the publication of Ref. \citenum{closs:1986:jcp:inverted}, Closs further investigated triplet energy transfer (TET) rates for the same set of molecules and gained further insight into nonadiabatic energy flow\cite{closs1988EET,closs1989dexter}. For more details on the relationship between ET and TET, please see the companion paper of this work (henceforward, Paper II)\cite{derosa20251dtet}.  For the purposes of this article, it must be emphasized that, in understanding their original work on TET \emph{via} cyclohexane and decalin bridges, Closs and co-workers were careful to distinguish different rates for different donor/acceptor attachments: \emph{i.e.} between both equatorially (`ee'),
equatorially and axially (`ea'), or axially and equatorially (`ae'). In general, as found in Ref. \citenum{subotnik2010closs} and as we will detail below, many simple features in the `ee' systems appear much more complicated for `ea' and `ae' systems; again, for more details see Paper II\cite{derosa20251dtet}.

\begin{figure*}
    \centering
    \includegraphics[width=\textwidth]{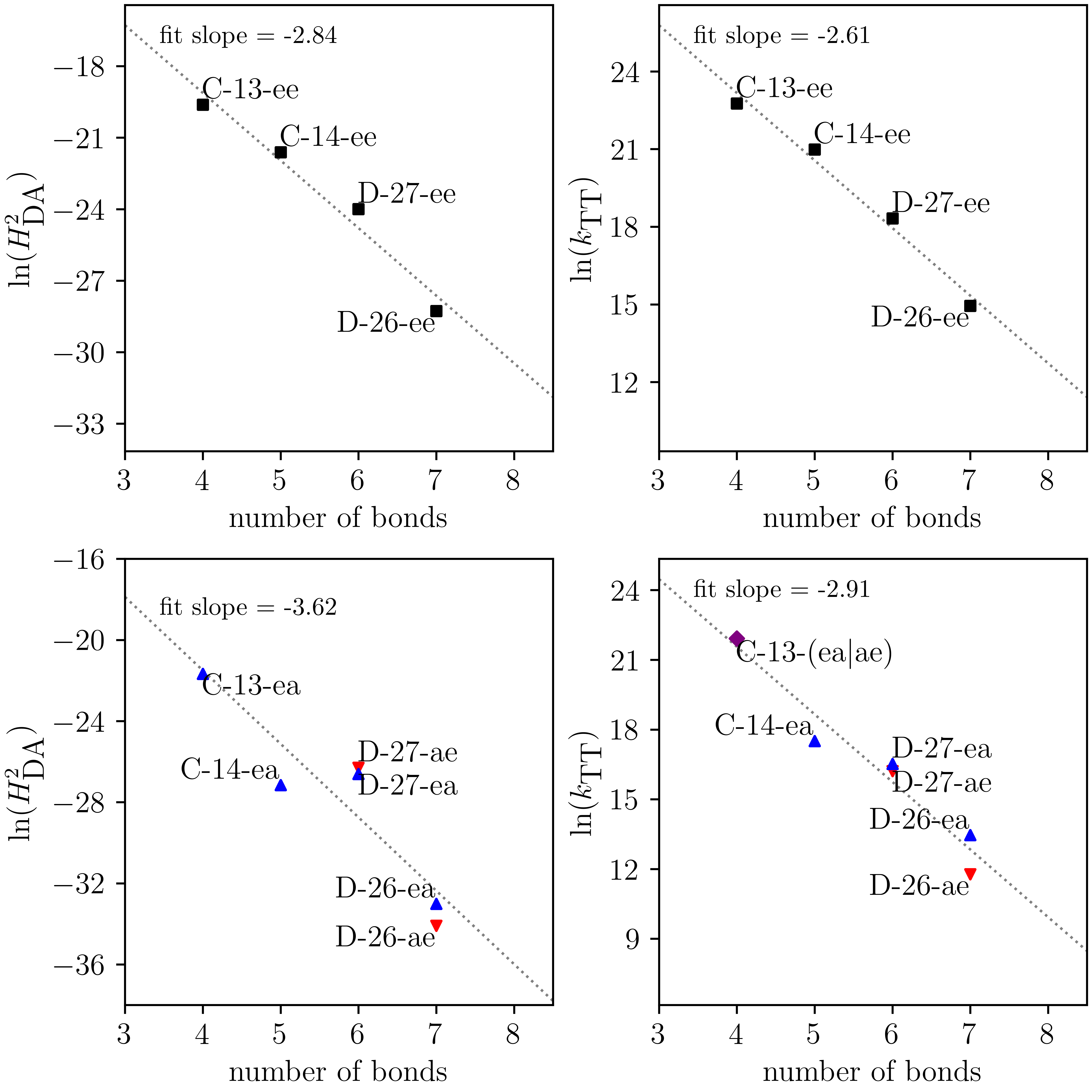}
    \caption{\label{fig:exptheory}Comparison between
    left: theoretically predicted couplings (ref.~\citenum{subotnik2010closs} and this work, `ae' systems) and
    right: experimentally determined rates (refs.~\citenum{closs1988EET,closs1989dexter}).
    Slopes to linear fits give inferred values for $\alpha$ in \refeq{eq:coupling-bonds}.
    The upper panels show the `ee' systems (black squares), which have excellent agreement between the the calculated couplings and measured rates.
    The lower panels show the `ae', red triangle-down, and `ea', blue triangle-up, systems.
    Note that Closs and co-workers did not identify whether they had the `ae' or `ea' variant of the C-13 bridge (purple diamond).
    For the `ae' and `ea' systems, the correlation between the rates and number of bonds is weaker than for the 'ee' systems; note the log scale on the y-axis.
    }
\end{figure*}

As mentioned above, one of the most appealing qualities of the Closs molecule is the ease with which one can simulate dynamics. 
To that end, already 15 years ago, our research group studied TET for the Closs systems using simple TD-DFT theory, together with
 Boys\cite{foster1960canonical} and  BoysOV localized diabatization \cite{subotnik2008constructing,subotnik2009initial} (see Eqs. \ref{eq:FU-boys} and \ref{eq:FU-boysOV} below).  For the `ee' molecules, we found perfectly quantitative data for theory vs. experiment.\cite{subotnik2010closs}. That being said however, the agreement for the `ae' and `ea' data was acceptable, but not amazing.
To emphasize the point, a graphical demonstration of the experimental and theoretical results from Refs. \citenum{closs1988EET}, \citenum{closs1989dexter} and \citenum{subotnik2010closs} are given in \fig{fig:exptheory}. 
For donor and acceptor linked by equatorial sites on the bridge, there was a simple relationship between the rate and the number of bonds ($n$)separating the sites implying that the coupling fell off exponentially with $n$:
\begin{equation}\label{eq:coupling-bonds}
H_\textrm{DA} \propto \exp(-\alpha n)
\end{equation}
The upper panels of \fig{fig:exptheory} show excellent agreement between calculated and measured values of $\alpha$ for `ee' systems;
however, there are large deviations when the donor or acceptor was substituted at an axial site as shown in the lower panels.
For the smallest molecule, with a freely-rotating methyl bridge, Marcus theory is clearly not applicable, and we successfully simulated dynamics with surface hopping trajectories\cite{landry:2014:abinitio_closs}.

Figure \ref{fig:exptheory} forms the basis for the present manuscript. One can ask: why do the rates of the `ea' and `ae' series of molecules not reliably decay with donor-acceptor separation distance? 
Given that the `ee' series decays with such a clean exponential form, in agreement with Eq.~\ref{eq:coupling-bonds},  why is the `ea'/`ae' series so much more complicated?
Could it be that the `ee' molecules are more rigid and the `ea'/`ae' molecules are floppier? If so, one would presume two consequences.
First,   the dynamics of a floppy molecule cannot be reduced to simple harmonic oscillators so that  Marcus theory may not be applicable at all (i.e. if one assumes that the vibration is the reaction coordinate). 
Second, if the molecule is floppy and explores more configuration space than for a rigid molecule, one would also expect a violation of the Condon approximation, which is also observed in Ref. \citenum{subotnik2010closs}.
Using the geometries from Ref. \citenum{subotnik2010closs}, we find that the RMSD between donor and acceptor configuration for C-13-ea (RMSD 0.13 \r{A}) is, in fact, over twice as large as for C-13-ee molecule (RMSD 0.06 \r{A}),
which suggests the `ea' molecules are indeed floppier than the `ee' ones.  
Finally and most interestingly,  for reasons not explained in the original experimental paper (Ref. \citenum{closs1988EET}), molecules C-13-ea and C-13-ae are reported to have the same rate constant.
But, given that the barrier to interconversion (as visualized in Figure \ref{fig:c13ae_c13ea_interconv}) is typically 10 \unit{kcal\cdot mol^{-1}} which is comparable to the barriers for other molecules (see Table \ref{tab:barriers} below), one would expect to be able to resolve all  `ae' and `ea' species independently.
Moreover, there is no reason that the rate of TET be the same for both configurations. Thus, one must also wonder: did the original experiments correctly interpret the C-13-ae/ea data?

To answer these questions and tease out the differences between the `ee' and `ea'/`ae' molecules, we introduce a new sampling scheme, E-SHAKE, that allows us to apply electronic constraints to nuclear dynamics. 
Specifically, we constrain \emph{ab initio} molecular dynamics trajectories to sample the seam of intersection between two diabatic potential energy surfaces.
In so doing, we can gain a great deal of intuition regarding the nature of the nuclear fluctuations and the magnitudes of those fluctuations necessary to promote an electronic transition for a non-harmonic system. 
The remainder of our paper is organized as follows: Section~\ref{sec:theory} reviews the construction of localized diabats (\ref{sec:theory-diabats}) and their gradients (\ref{sec:theory-gradients}) and introduces E-SHAKE (\ref{sec:theory-eshake});
our results are presented and discussed in Sections~\ref{sec:results} and~\ref{sec:discussion}, respectively; in Section~\ref{sec:conclusion} we conclude and comment about the future possibilities for  E-SHAKE.

\section{Theory and Methods}\label{sec:theory}

\subsection{Constructing Localized Diabats}\label{sec:theory-diabats}
To compute the coupling element, $H_{\textrm{DA}}$, we require diabatic wavefunctions of constant character to represent the donor and acceptor states.
In general, diabats may be formed by rotating the adiabatic states into a new non-diagonal basis:
\begin{equation}
    \ket{\Xi_i} = \sum_{j=1}^{N_{\textrm{states}}} \left(\mat{U}\right)_{ji}\ket{\Phi_j} \quad i=1\ldots N_{\textrm{states}}
\end{equation}
Here, $\mat{U}$ is an adiabatic-to-diabatic transformation matrix\cite{pacher1988approximately,pacher1993adiabatic}. While there are no unique, global criteria for $\mat{U}$\cite{mead1982conditions}, for electron transfer\cite{subotnik2009initial}, one standard approach is to utilize the Boys criterion\cite{subotnik2008boys} and maximize:
\begin{equation}\label{eq:FU-boys}
    f_{\textrm{Boys}} (\mat{U}) = \sum_{i,j=1}^{N_{\textrm{states}}}
    \abs{\ev{\hat{\mu}}{\Xi_i} - \ev{\hat{\mu}}{\Xi_j}}^2,
\end{equation}
which produces diabats with maximally distinct dipole moments.
While the condition is inapplicable to energy excitation transfer, a very similar approach that is relevant to energy transfer is to separately localize the electronic excitations and de-excitations of the excited states--the so-called ``BoysOV''approach\cite{subotnik2008boys}.
In practice, we maximize a dipole operator separated into occupied and virtual components:
\begin{align}\label{eq:FU-boysOV}
    f_{\textrm{BoysOV}} (\mat{U}) = \sum_{i,j=1}^{N_{\textrm{states}}}
    &\abs{\ev{\hat{\mu}^{\textrm{occ}}}{\Xi_i} - \ev{\hat{\mu}^{\textrm{occ}}}{\Xi_j}}^2 + \nonumber \\
    &\abs{\ev{\hat{\mu}^{\textrm{virt}}}{\Xi_i} - \ev{\hat{\mu}^{\textrm{virt}}}{\Xi_j}}^2
\end{align}
While we used the Boys/BoysOV method in this work, there are other methods that have also been shown to produce localized diabats for ET and TET states,\cite{cave1996generalization, subotnik2009initial,voityuk2002fragment,hsu2008characterization} for example the fragment charge difference method explicitly calculates net fragment charges on donor and acceptor to determine the adiabatic-to-diabatic transformation\cite{voityuk2002fragment} and this can be extended to spin and energy transfer as well.\cite{hsu2008characterization}
Having constructed an appropriate $\mat{U}$, we can rotate from the adiabatic basis into a diabatic one and directly arrive at the coupling, $H_\textrm{DA}$:
\begin{equation}
    \mat{U}^{\ddagger}
    \begin{bmatrix}
        \textrm{E}_1 & 0 \\
        0 & \textrm{E}_2
    \end{bmatrix} 
    =
    \begin{bmatrix}
        E_\textrm{D} & H_{\textrm{DA}} \\
        H_{\textrm{DA}} & E_\textrm{A}
    \end{bmatrix}
\end{equation}
where $E_1$ and $E_2$ are the energies of two adiabatic states and $E_\textrm{D}$ and $E_\textrm{A}$ are the energies associated with the donor and acceptor diabats, respectively.

\subsection{Analytic gradients}\label{sec:theory-gradients}
Though there has been recent progress in constructing exact gradients of diabatic states\cite{glover2023diabatgradients,athavale2025evaluating}, we construct approximate gradients\cite{fatehi2013diabatderivatives} by invoking the strictly diabatic approximation, which assumes that the derivative coupling between diabatic states is strictly zero:
\begin{equation}
    \mel{\Xi_i}{\vec{\nabla}_{R}}{\Xi_j} = 0 
\end{equation}
Note that the use of an approximate \emph{diabatic} gradient does not strongly affect the dynamics insofar as, despite any inherent systematic error, the constraint algorithm below will insist that we we remain on the diabatic seam (which can be checked numerically without a gradient). See below.

\subsection{E-SHAKE: Sampling a Diabatic Seam}\label{sec:theory-eshake}
Previous efforts to explore the seam-space of an electronic intersection have used biasing potentials\cite{memes}, enhanced sampling with modifications to the electronic Hamiltonian\cite{metafalcon} or a constrained extension to the nudged elastic band method\cite{mori2013nebci}. However, none of these methods permit the exhaustive exploration of seams in a generic fashion. E-SHAKE, which we now present, allows sampling a diabatic or adiabatic seam far from a minimum energy crossing point and requires only gradients of the intersecting surfaces.

To impose an electronic constraint on the nuclear coordinates sampled during \emph{ab initio} molecular dynamics, the equations of motion must be modified.
Suppose we demand that the donor and acceptor states have the same energy; this yields a single constraint on the configuration of the nuclei:
\begin{equation}\label{eq:constraint-egap}
    E_{\textrm{D}}(\vec{R})=E_{\textrm{A}}(\vec{R}),
\end{equation}
Thus, for a system with $3N$ degrees of freedom, the seam between two diabatic energy surfaces is a reduced-dimensionality configuration space of dimension $3N-1$.
To sample the distribution of configurations in this reduced space, we implement an \emph{electronic} variant of the SHAKE and RATTLE algorithms\cite{andersen1983RattleVelocityVersion}, thus ``E-SHAKE.''
In accordance with Eq. \refeq{eq:constraint-egap},  the present work samples a fixed donor-acceptor energy gap, as implemented in INAQS\cite{inaqs:1}/Gromacs\cite{gromacs45}.
However, as implemented, E-SHAKE can impose arbitrary electronic constraints on the nuclear dynamics, including sampling the seam between two spin diabats\cite{derosa2024seam}.

The original SHAKE and RATTLE alorithmgs are implemented using Lagrange multipliers to compute forces of constraints analytically such that the dynamics satisfies a geometric equality, \emph{e.g.} fixing a bond length or angle.
For E-SHAKE however, we do not have an analytic form of the constraint, which is computed via \emph{ab initio} means, thus we must solve the constraints iteratively. 
The goal is to propagate molecular dynamics along the relevant potential energy surface $V(\vec{R})$ according to Eq. \refeq{eq:fma} while simultaneously respecting the two constraints as given by Eq. \refeq{eq:const:shake} and \refeq{eq:const:rattle}.
\begin{align}
    \ddot{\vec{R}} &= -\mat{M}^{-1} \cdot {\left. \vec{\nabla} V \right|}_{\vec{R}} \label{eq:fma} \\
    (\mat{M})_{i\mu,j\nu} &= \delta_{i,j}m_i  \quad i,j=1\ldots N ; \mu,\nu\in\{x,y,z\}\nonumber
\end{align}
\begin{align}
0 = \sigma(t) &=\sigma(\vec{R}(t)) \label{eq:const:shake}\\
0 = \dot{\sigma}(t) &=\dot{\vec{R}}(t) \cdot {\left. \vec{\nabla} \sigma \right|}_{\vec{R}(t)} .\label{eq:const:rattle}
\end{align}
Note that $\{m_i\}$ is the mass of the atom $i$.
\refeq{eq:const:shake} imposes a constraint on the nuclear coordinates and \refeq{eq:const:rattle} ensures that the velocity at each step is also directed within the manifold of constraint.
For arbitrary $\sigma$, Eqs.~\ref{eq:const:shake} and~\ref{eq:const:rattle} are generalizations of the SHAKE and RATTLE conditions, respectively.
For our purposes, we take the constraint
\begin{equation}
    \sigma(\vec{R}) = E_{\textrm{D}}(\vec{R}) - E_{\textrm{A}}(\vec{R}),
\end{equation}
to encode the difference between the energy of the donor, $E_{\textrm{D}}$, and the acceptor, $E_{\textrm{A}}$, at the geometry $\vec{R}$.

We use the velocity variant of the Verlet integrator for a number of reasons: computational efficiency, long term stability, and allowing for thermostatting the dynamics.\cite{verlet,vverlet}.
Suppose we have a system evolving according to the differential equation in \refeq{eq:fma}, and we hope to enforce constraints of the form of Eqs.~\ref{eq:const:shake} and \ref{eq:const:rattle} for all time, $t$. 
The unconstrained velocity-Verlet equations are given as follows:
\begin{align}
    \vec{R}(t+h) &= \vec{R}(t) + h \dot{\vec{R}}(t) + \frac{h^2}{2} \mat{M}^{-1} \vec{F}(t) \label{eq:verlet:r} \\
    \dot{\vec{R}} (t+h) &= \dot{\vec{R}}(t) + \frac{h}{2} \mat{M}^{-1} [\vec{F}(t) + \vec{F}(t+h)] \label{eq:verlet:v}
\end{align}
where $\vec{F}(t)= -{\left. \vec{\nabla}V\right|}_{\vec{R}(t)}$ and $h$ is the time step.
To satisfy the constraints in Eqs.~\ref{eq:const:shake} and~\ref{eq:const:rattle}, we introduce two Lagrange multipliers
$\lambda_{RR}$ and $\lambda_{RV}$ and write the resulting constrained equations of motion as:
\begin{align}
    \vec{R}(t+h) &= \vec{R}(t) + h \dot{\vec{R}}(t) + \frac{h^2}{2} \mat{M}^{-1} [\vec{F}(t) + \lambda_{RR} \vec{G}(t)] \label{eq:rattle:r} \\
    \dot{\vec{R}} (t+h) &= \dot{\vec{R}}(t) + \frac{h}{2} \mat{M}^{-1} [\vec{F}(t) + \lambda_{RR} \vec{G}(t) + 
    \vec{F}(t+h) + \lambda_{RV} \vec{G}(t + h)]\label{eq:rattle:v}
\end{align}
where $\vec{G}(t) = -{\left. \vec{\nabla}\sigma \right|}_{\vec{R}(t)}$.
Defining
\begin{align}
    g &\equiv h \lambda_{RR} , \quad
    k  \equiv h \lambda_{RV} , \\
    \vec{Q} &\equiv \dot{\vec{R}}(t) + \frac{1}{2} \mat{M}^{-1} \cdot \left[ h\vec{F}(t) + g\vec{G}(t)\right],
\end{align}
Eqs~\ref{eq:rattle:r} and \ref{eq:rattle:v} can be rewritten as:
\begin{align}
    \vec{R}(t+h) &= \vec{R}(t) + h \vec{Q} \nonumber \\
    \dot{\vec{R}}(t + h) &= 
    \vec{Q} + \frac{1}{2} \mat{M}^{-1}\cdot \left[ h \vec{F}(t + h) + k\vec{G}(t+h)\right]\label{eq:rattle_simplified}
\end{align}
Our task is now to solve for $g$ and $k$ such that $\abs{\sigma(\vec{R}(t))}<\epsilon$ for some choice of threshold of $\epsilon$. For the work reported here, we enforce a threshold of $\epsilon = 4\times 10^{-5} \,\textrm {a.u.} \approx 1 \,\textrm{meV}$.
The solution for $g$ may be obtained by solving:
\begin{equation}
    \sigma \left( \tilde{\vec{R}}(t+h) + g \frac{h}{2}\mat{M}^{-1}\cdot\vec{G}(t)\right) = 0 \label{eq:const:g}
\end{equation}
where $\tilde{\vec{R}}(t+h)$ corresponds to an unconstrained step resulting from \refeq{eq:verlet:r}.
While SHAKE was originally developed for geometric, analytic constraints, solving \refeq{eq:const:g} is effectively a 1-dimensionally root search and can be peformed numerically. \cite{vale_footnote}
Our implementation uses bisection with assumption of linearity and a single Newton-Raphson step for bracketing the initial guess\cite{press2007}.

After obtaining a solution for $g$, \refeq{eq:const:rattle} with the velocity from \refeq{eq:rattle:v} can be solved directly for k: 
\begin{equation}
    {\left[ \tilde{\dot{\vec{R}}}(t + h) + \frac{k}{2} \mat{M}^{-1} \vec{G}(t + h)\right]}^{\textrm{T}} \cdot \vec{G}(t+h) = 0
\end{equation}
Here, we have defined the partially unconstrained velocity (recall that $\vec{Q}$ contains $g$):
\begin{equation}
    \tilde{\dot{\vec{R}}}(t + h) = \vec{Q} + \frac{h}{2} \mat{M}^{-1}\cdot \vec{F}(t + h).
\end{equation}
The solutions for $k$ are trivial:
\begin{equation}\label{eq:k_solution}
    k = 2 \frac{\tilde{\dot{\vec{R}}}^{\textrm{T}}(t + h)
    \cdot \vec{G}(t + h)}{\vec{G}^{\textrm{T}}(t + h) \cdot \mat{M}^{-1} \cdot \vec{G}(t + h)}
\end{equation}
and satisfy \refeq{eq:const:rattle} to machine precision. With the novel E-SHAKE algorithm in hand, it is important to take note that the added computational cost is simply two additional diabatic gradient calculations at each time step.

\subsection{Computational Details}\label{sec:theory-computational_details}
All electronic structure calculations were performed using configuration interaction singles (CIS) in the 6-31G* basis set in Q-Chem 6.0\cite{qchem}. For E-SHAKE, geometries  were initialized in the vicinity of a minimum energy crossing point between the first two triplet surfaces and propagated along the $T_1$ surface under the constraint that the donor-acceptor diabatic energy gap be less than $4\times10^{-5}$ Hartree ($\approx1$ meV), which was enforced with the E-SHAKE method described above in section \ref{sec:theory-eshake}. Seam sampling dynamics were thermostatted using a stochastic modified Berendsen method\cite{bussi2007thermostat} with a time constant, $\tau=10\,\textrm{fs}$.

\section{Results }\label{sec:results}
\begin{figure}
    \centering
    \includegraphics[width=\columnwidth]{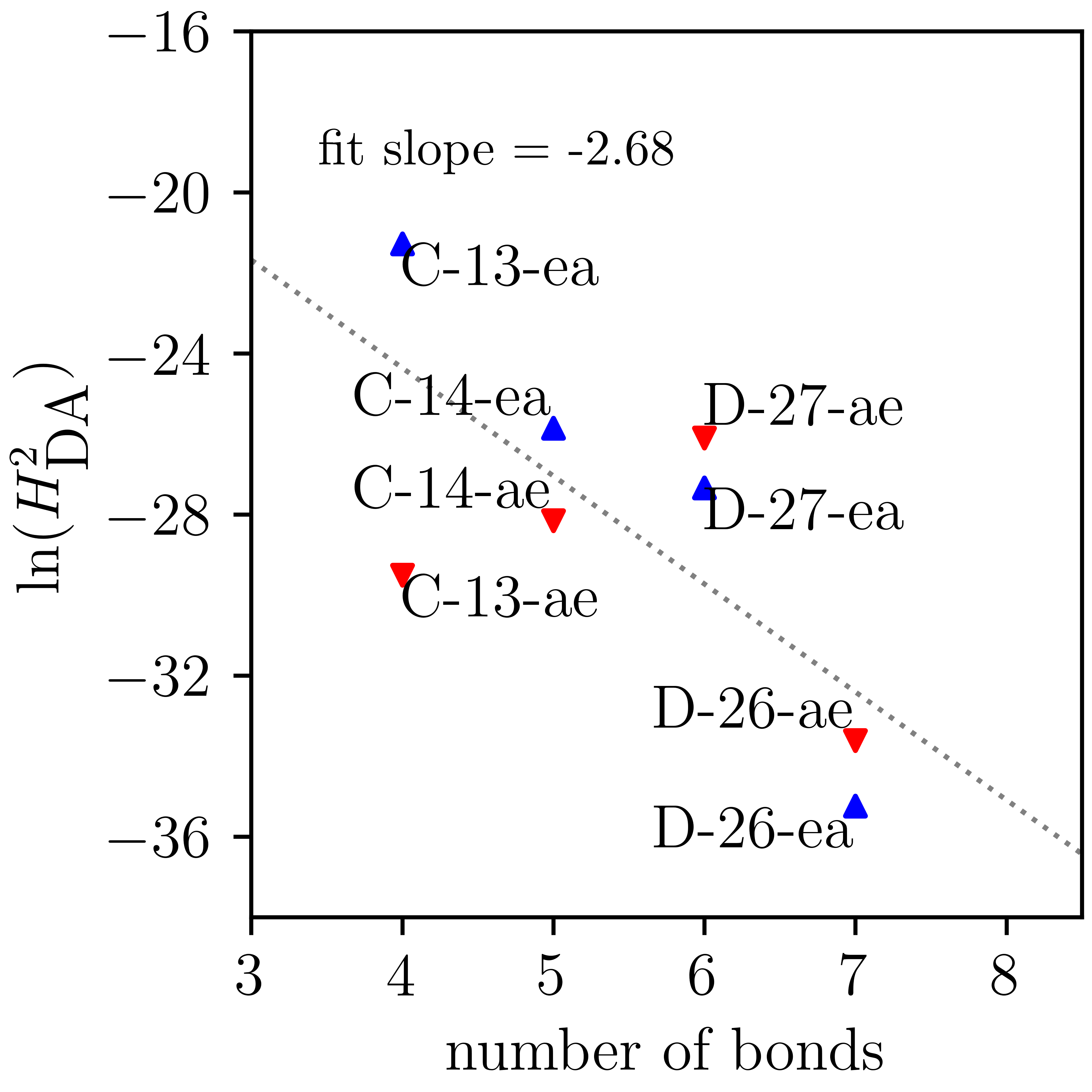}
    \caption{\label{fig:exptheory-ae} As in \fig{fig:exptheory}, a comparison between computed diabatic coupling and number of bonds for `ae' and `ea' systems. Couplings are computed in the vicinity of the $\textrm{T}_1$/$\textrm{T}_2$ crossing and are largely the same as at the ground state minimum. However, here we also plot the `ae' variants of the C-13 and C-14 bridges. While C-14-ae is relatively similar to C-14-ea, C-13-ae represents a particularly striking outlier that we will return to below.
    }
\end{figure}

To begin our analysis, we recomputed the donor-acceptor couplings for each bridge and donor-acceptor attachment `ea'/`ae' scheme from \fig{fig:exptheory}  at a configuration near the $\textrm{T}_1$/$\textrm{T}_2$ crossing.
Figure~\ref{fig:exptheory-ae} summarizes our results. For the decalin bridged systems (D), we find the `ae' and `ea' variants are similar.  We find that C-14-ae and -ea also show similar couplings. However, most strikingly, we find that C-13-ae has a computed coupling two orders of magnitude smaller than C-13-ea.

\begin{figure}
    \centering
    \includegraphics[width=\columnwidth]{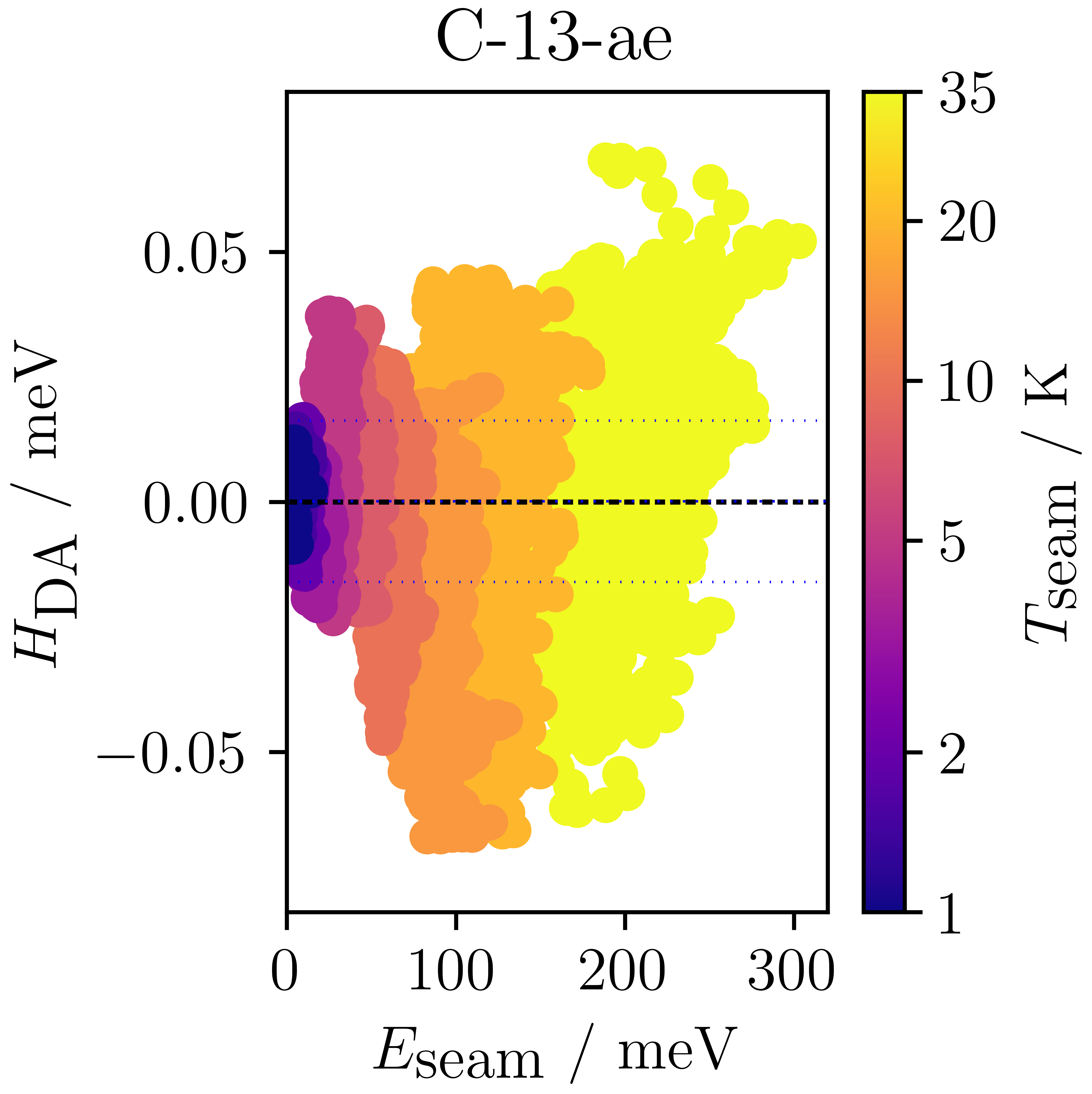}
    \caption{\label{fig:seam-13ae}Coupling between donor and acceptor diabats as a function of energy within the seam above the minimum for \ce{C-13-ae}. Note that for this molecule, the diabatic coupling is centered around zero, indicating the presence of a CI.}
\end{figure}

\begin{figure}
    \centering
    \includegraphics[width=\columnwidth]{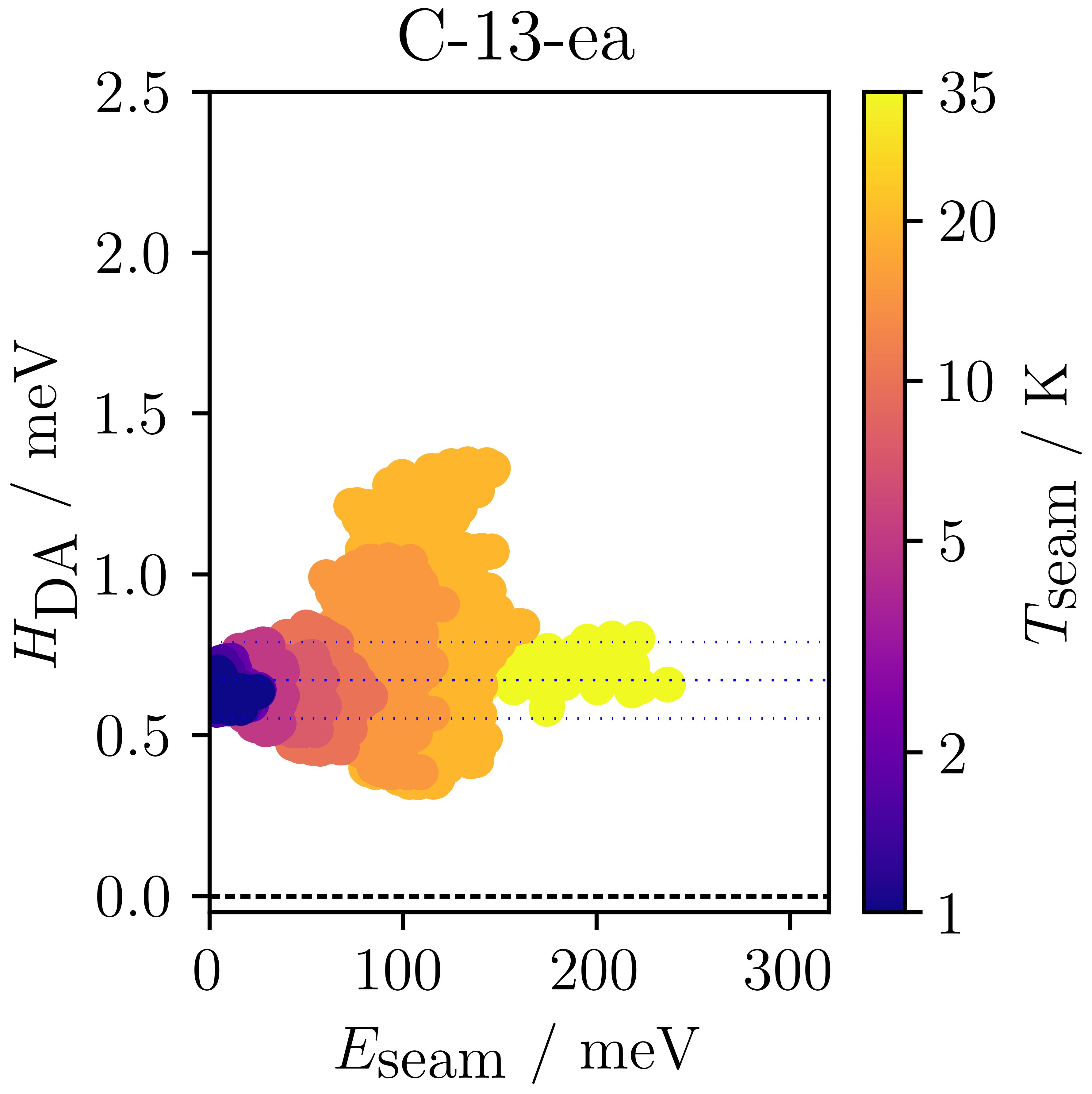}
    \caption{\label{fig:seam-13ea}Coupling between donor and acceptor diabats as a function of energy within the seam above the minimum for \ce{C-13-ea}.  Note that for this molecule, the diabatic coupling is not centered at zero and is reasonably constant, reflecting a reasonable adherence to the Condon approximation.}
\end{figure}

\begin{figure}
    \centering
    \includegraphics[width=\columnwidth]{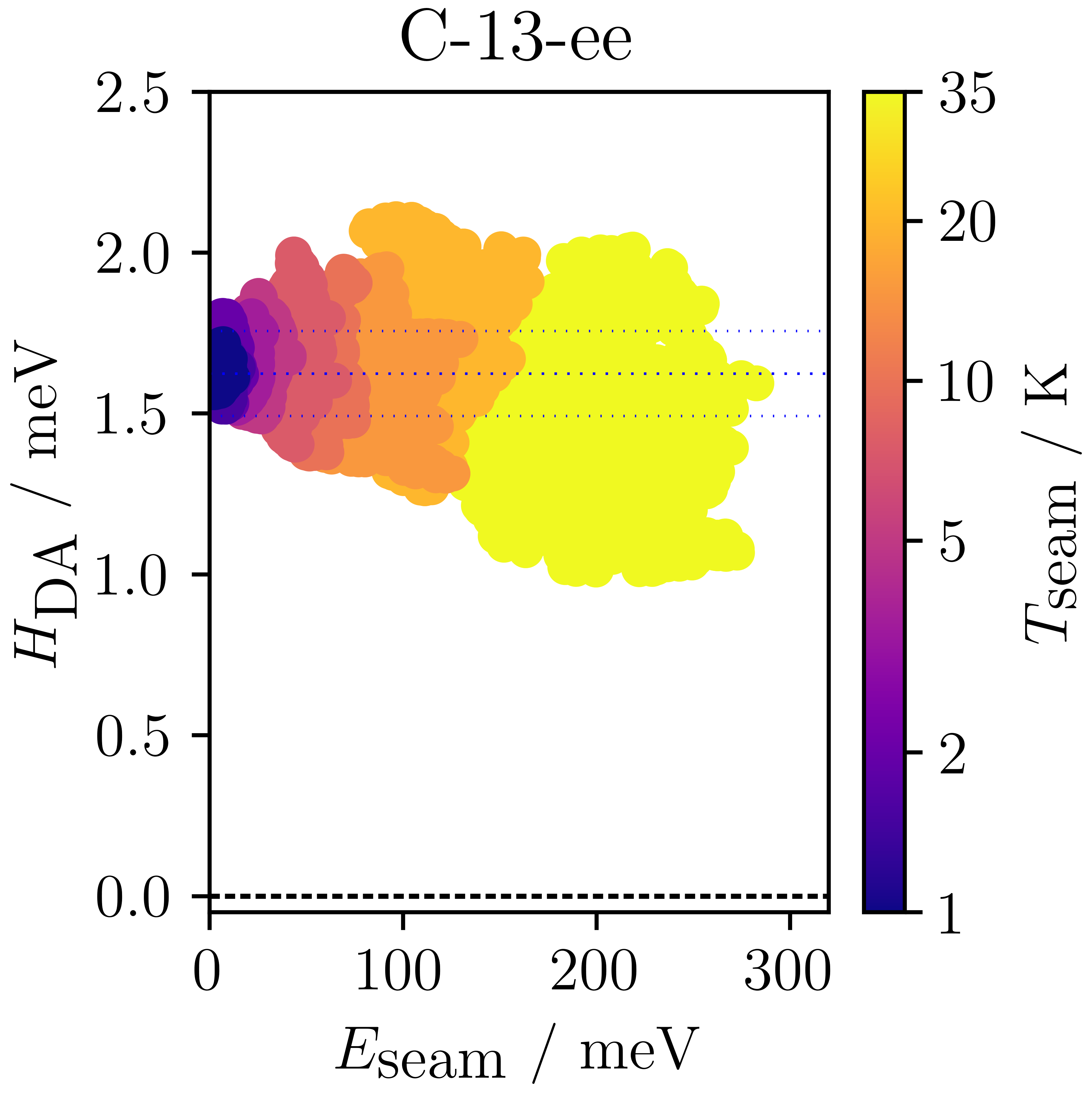}
    \caption{\label{fig:seam-13ee}Coupling between donor and acceptor diabats as a function of energy within the seam above the minimum for \ce{C-13-ee}. Note that for this molecule, the diabatic coupling is not centered at zero and is reasonably constant, reflecting a reasonable adherence to the Condon approximation.}
\end{figure}

The data in Figures~\ref{fig:exptheory} and \ref{fig:exptheory-ae} represents only one geometry per molecule. 
To better investigate the relevant dynamics and the discrepancy between variants of the C-13 bridge, we next turned to seam sampling, propagating constrained dynamics along the $\textrm{T}_1$ surface, such that at all times the gap between donor and acceptor diabats is less than 1~\unit{meV}.
While running the seam sampling, we  monitored the donor-acceptor coupling, $H_{\textrm{DA}}$, as a function of nuclear coordinate.
In Figures~\ref{fig:seam-13ae}, \ref{fig:seam-13ea}, and \ref{fig:seam-13ee} we show the donor-acceptor coupling  ($H_{\textrm{DA}}$) as a function of energy \emph{within} the donor-acceptor seam for the molecules C-13-ea, C-13-ae, and C-13-ee.
The zero of the abscissa, $E_{\textrm{seam}}$, is the minimum $\textrm{T}_1$ energy encountered: higher energies are higher above the minimum energy crossing in the diabatic seam.
Color indicates the the sampling ``temperature,'' $T_{\textrm{seam}}$---higher temperatures explore higher energy regions of the seam, but the temperature should not be thought of as physical.

Aside from the relative differences in scale between the the coupling for C-13-ae and C-13-ea, the most striking difference is passage of the coupling through zero.
For both C-13-ea and C-13-ee (and indeed for all the variants of the C-14 bridge), the coupling is relatively tightly distributed about a finite mean value (dashed lines on each figure give the mean $\pm1$ standard deviation).
This consistency is the essence of the Condon approximation: the coupling is relatively constant with geometry.
In stark contrast, however, the couplings for C-13-ae are distributed about zero.
In fact, the magnitude of $H_{\textrm{DA}}$ spans two orders of magnitude within the seam, indicating a system that simply does not obey the Condon approximation.
Furthermore, in the limit that donor and acceptor energies are the same and the coupling goes to zero, we move from a diabatic seam to a conical intersection (CI).
\begin{equation}
\lim_{\substack{H_{\textrm{DA}} \to 0 \\ E_\textrm{D} \to E_\textrm{A}=E}}
    \begin{bmatrix}
        E_\textrm{D} & H_{\textrm{DA}} \\
        H_{\textrm{DA}} & E_\textrm{A}
    \end{bmatrix}
    =
    \begin{bmatrix}
        E & 0 \\
        0 & E
    \end{bmatrix}
\end{equation}

The end result is that we predict that, if one goes back and re-measures these results, we will find that C-13-ea and C-13-ae will have vastly different TET rates; there should be a very slow C-13-ae rate scale that was not observed at the time\cite{closs1988EET}.  (Note that, while the presence of a CI would tend to speed up relaxation downwards from an excited state and act like a funnel, one expects that a CI should delay thermal, activated energy transfer as a trajectory would need to reach a higher energy barrier in order to experience a diabatic coupling along the seam.)  This difference in rate constant can be attributed to a failure of the Condon approximation and the inapplicability of using Marcus's expression (Eq. \refeq{eq:marcus-eet}) in the vicinity of a conical intersection. Indeed, this data should remind us of the dangers of assuming that all diabatic couplings fall off as a function of bond distance (like \refeq{eq:coupling-bonds}) as usually assumed in the literature; several researchers have even predicted that electron transfer rates have little to do with bonding structure but only total length.\cite{moser1992nature}

Finally, as an important sanity check, we note that  we do not observe any kinetic reason such that the `ae' and `ea' molecules will thermally interconvert for the C-13 bridge faster than the the other bridges.  According to our calculations, the ground-state interconversion are relatively similar for all systems in \fig{fig:exptheory}: 9--11 \unit{kcal\cdot mol^{-1}} (see Table~\ref{tab:barriers}).
\begin{table}[]
    \centering
    \begin{tabular}{l c}
        bridge & barrier / \unit{kcal\cdot mol^{-1}} \\
        \hline
        C-13 & 9.2  \\
        C-14 & 9.4  \\
        D-26 & 9.7  \\
        D-27 & 11.1
    \end{tabular}
    \caption{\label{tab:barriers} Ground state barrier for the interconversion of the `ae' and `ea' variants for each bridge. Barriers were found using the freezing string method\cite{behn2011fsm} followed by transition state search.}
\end{table}

\section{Discussion}\label{sec:discussion}
\begin{figure}
    \centering
    \includegraphics[width=\columnwidth]{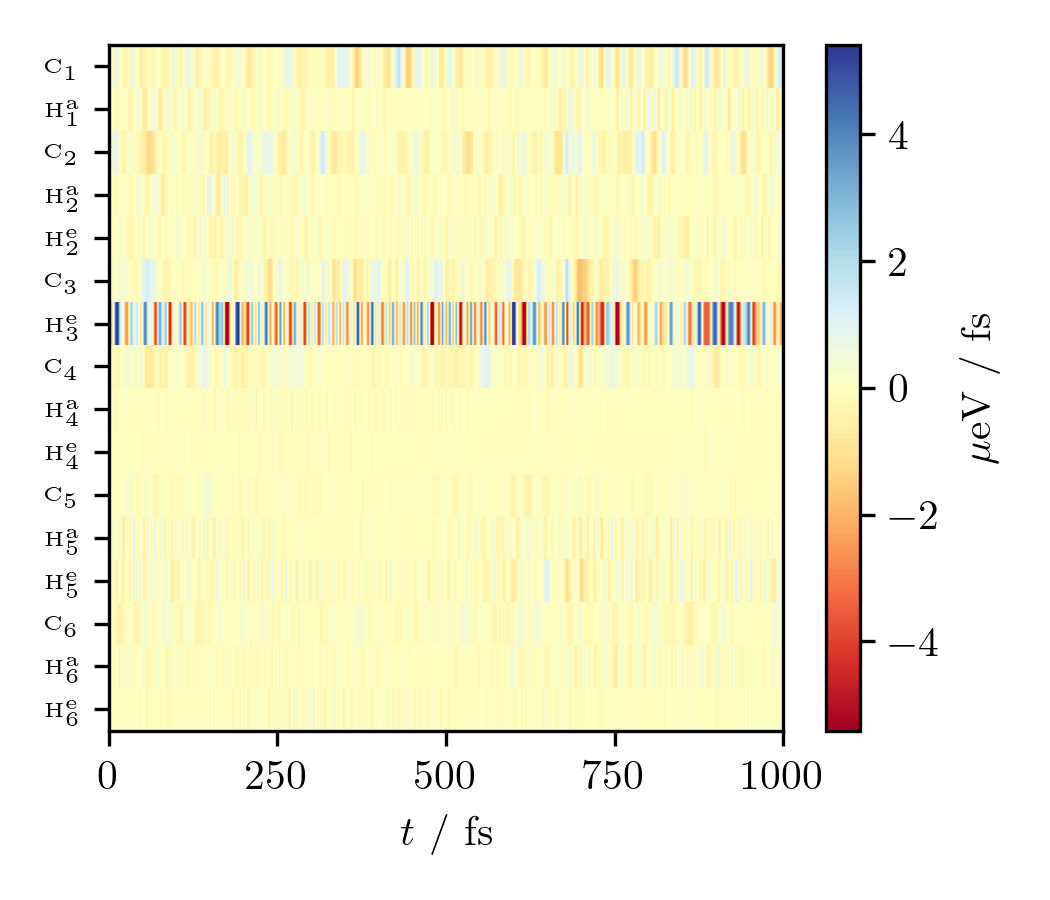}
    \caption{\label{fig:dvdt} Per-atom contribution to change in the donor-acceptor coupling (\refeq{eq:dVadt}) as a function of time in the seam for the C-13-ae complex. Sampling temperature, $T_{\textrm{seam}}=35\,\textrm{K}$. The vertical axis lists the bridge atoms in terms of their position on on the cyclohexane ring; hydrogens are additionally labeled to indicate whether they are in (a)xial or (e)quatorial positions. One atom stands out as primarily contributing to the fluctuations in the coupling: the hydrogen below the donor benzaldehyde.
    }
\end{figure}

In short, we have made a prediction about a new rate that can be measured; there should be another (slower) rate for C-13-ae TET that can be found. Furthermore, we would also like to suggest a possible isotopic substitution that would confirm our prediction of a CI.
Noting that the C-13-ae diabatic seam is quite floppy, the per-atom contribution to the change in the coupling can be determined in terms of the total gradient and the velocity by summing over the spatial degrees of freedom:
\begin{equation}
    \label{eq:dVadt}
    {\frac{dV}{dt}}^{(i)} = \sum_{\mu=x,y,z} \frac{dV}{dR_{i\mu}}\frac{dR_{i\mu}}{dt} .
\end{equation}
As shown in \fig{fig:dvdt}, the fluctuations are dominated by the motion of a single hydrogen atom in the equatorial site below the donor ($\mathrm{H_3^e}$).
Analysis of the attachment and detachment densities\cite{mhg1995attachdetach} for the donor and acceptor wavefunctions confirms fluctuations in the donor/acceptor overlap to be  driven by motion of the same hydrogen.
We performed a similar decomposition over normal modes, but invocation of normal modes only made the fluctuations more delocalized and was otherwise uninformative.

Assuming that dynamics pass through a CI, then it follows that substituting D for $\mathrm{H_3^e}$ should have a large isotopic effect for C-13-ae (but not C-13-ea). To make a quantitative argument for the exact size of such effect would require a more detailed potential energy surface and more elaborate Fermi golden rule or master equation calculations. But at the very least, we expect that when going through a CI, we would observe a much faster rate with D as opposed to H. 

Before concluding, there is one final question that must be asked: Is the fact that C-13-ae passes through a conical intersection an anomaly, or are conical intersections ubiquitous for TET? On that front, we note that there has been a great deal of discussion about the ubiquity of CIs---in the early history of work on radiationless transitions, CIs were considered rare indeed; later, CIs were found everywhere\cite{yarkony1996diabolical}. In some senses, the question above is then ill posed: CIs are everywhere and hard to detect\cite{yarkony1996diabolical,manaa1993intersection,bearpark1994direct}, but their effect is strongest only when they lie very low in energy and working out exactly when that occurs must depend molecule by molecule. Within the context of ET, however, the thought to this point has been that ultrafast CT is often mediated through a CI,\cite{domcke2012role} though measuring the rate of ultrafast CT events has been proven difficult and has only seen success in cases of photoexcited-induced charge transfer.\cite{fuss2002ultrafast,rettig1986charge,rettig1992photoinduced} In any event, this paper highlights one example of a CI for TET, and in a companion paper\cite{derosa20251dtet}, we will argue that the Condon approximation is more likely violated for TET than for ET. More generally, as to the ubiquity of low-lying CIs for TET processes,  we do not yet have a strong feeling and a great deal of future simulation will be necessary; after all, the existence of a CI requires geometries with both zero coupling and zero energy gap, and so far we have analyzed only the diabatic coupling in depth.

\section{Conclusion}\label{sec:conclusion}

The findings of our work indicate that non-Condon effects can be paramount for TET processes.
To that end, we have developed a seam sampling  technique that directly simulates  accessible geometries for molecules undergoing an electronic transition. We expect E-SHAKE will be applicable to many future systems of interest to gain more and more intuition about what molecular motion modulates a nonadiabatic crossing.
We note that, for case studied here, the seam is likely at or near the transition state, but the present algorithm can also be used for a crossing that is not exactly at the transition state (e.g. see Ref. \citenum{derosa2024seam}).
Lastly, here seam sampling is applied to a molecule, but it could be more generally applied to a solution and inform us about how solvent fluctuations affect the diabatic coupling or the landscape of the diabatic seam space more generally.
Interestingly, if combined with  analytic theories of how solvent can affect the diabatic coupling,\cite{hsu2009electronic,krueger1998calculation,hsu2001role}
the E-SHAKE approach would seem to offer a fresh approach toward understanding electron transfer in concentrated electrolyte solutions (i.e. well beyond the simple two-state Marcus picture). 

Finally, in this work we have focused on the anomalous Closs molecules as they relate to the exponential decay of the rate as predicted by Marcus theory. While the `ee' series has been known to reliably decay with bridge length, the agreement between theory and experiment is much weaker for the `ea'/`ae' molecules. We have shown that the C-13-ae and -ea molecule breaks a number of assumptions necessary for the Marcus theory rate to apply: (1) the `ea' molecules are less rigid than the `ee' ones, (2) the Condon approximation is not upheld as couplings are highly oscillatory, and, (3) at least for the C-13-ae molecule, there is an energetically accessible CI.
Thus, our prediction is that, with improvements in ultrafast spectroscopy over the last 20 years\cite{krausz2009attosecond}, if properly synthesized, one will find one new (very slow) Closs rate that has not yet been quantified, highlighting just how much we still have to learn about electron and energy transfer after all these years.

\begin{acknowledgement}
This work was supported by the U.S. Air Force Office of Scientific Research (USAFOSR)
under Grant No. FA9550-23-1-0368.
We thank the DoD High Performance Computing Modernization Program for computer time.
This material is based upon work supported by the National Science Foundation Graduate Research Fellowship under Grant No. DGE-2236662.
\end{acknowledgement}

\bibliography{main}

\providecommand{\latin}[1]{#1}
\makeatletter
\providecommand{\doi}
  {\begingroup\let\do\@makeother\dospecials
  \catcode`\{=1 \catcode`\}=2 \doi@aux}
\providecommand{\doi@aux}[1]{\endgroup\texttt{#1}}
\makeatother
\providecommand*\mcitethebibliography{\thebibliography}
\csname @ifundefined\endcsname{endmcitethebibliography}
  {\let\endmcitethebibliography\endthebibliography}{}
\begin{mcitethebibliography}{60}
\providecommand*\natexlab[1]{#1}
\providecommand*\mciteSetBstSublistMode[1]{}
\providecommand*\mciteSetBstMaxWidthForm[2]{}
\providecommand*\mciteBstWouldAddEndPuncttrue
  {\def\EndOfBibitem{\unskip.}}
\providecommand*\mciteBstWouldAddEndPunctfalse
  {\let\EndOfBibitem\relax}
\providecommand*\mciteSetBstMidEndSepPunct[3]{}
\providecommand*\mciteSetBstSublistLabelBeginEnd[3]{}
\providecommand*\EndOfBibitem{}
\mciteSetBstSublistMode{f}
\mciteSetBstMaxWidthForm{subitem}{(\alph{mcitesubitemcount})}
\mciteSetBstSublistLabelBeginEnd
  {\mcitemaxwidthsubitemform\space}
  {\relax}
  {\relax}

\bibitem[Nitzan(2006)]{nitzanCDCP}
Nitzan,~A. \emph{Chemical dynamics in condensed phases: relaxation, transfer,
  and reactions in condensed molecular systems}; Oxford university press,
  2006\relax
\mciteBstWouldAddEndPuncttrue
\mciteSetBstMidEndSepPunct{\mcitedefaultmidpunct}
{\mcitedefaultendpunct}{\mcitedefaultseppunct}\relax
\EndOfBibitem
\bibitem[Medvedev and Stuchebrukhov(1997)Medvedev, and
  Stuchebrukhov]{medvedev1997inelastic}
Medvedev,~E.~S.; Stuchebrukhov,~A.~A. Inelastic tunneling in long-distance
  biological electron transfer reactions. \emph{The Journal of chemical
  physics} \textbf{1997}, \emph{107}, 3821--3831\relax
\mciteBstWouldAddEndPuncttrue
\mciteSetBstMidEndSepPunct{\mcitedefaultmidpunct}
{\mcitedefaultendpunct}{\mcitedefaultseppunct}\relax
\EndOfBibitem
\bibitem[Jang and Newton(2005)Jang, and Newton]{jang2005theory}
Jang,~S.; Newton,~M.~D. Theory of torsional non-Condon electron transfer: A
  generalized spin-boson Hamiltonian and its nonadiabatic limit solution.
  \emph{The Journal of chemical physics} \textbf{2005}, \emph{122},
  024501\relax
\mciteBstWouldAddEndPuncttrue
\mciteSetBstMidEndSepPunct{\mcitedefaultmidpunct}
{\mcitedefaultendpunct}{\mcitedefaultseppunct}\relax
\EndOfBibitem
\bibitem[Tully(1990)]{tully1990molecular}
Tully,~J.~C. Molecular dynamics with electronic transitions. \emph{The Journal
  of Chemical Physics} \textbf{1990}, \emph{93}, 1061--1071\relax
\mciteBstWouldAddEndPuncttrue
\mciteSetBstMidEndSepPunct{\mcitedefaultmidpunct}
{\mcitedefaultendpunct}{\mcitedefaultseppunct}\relax
\EndOfBibitem
\bibitem[Ben-Nun \latin{et~al.}(2000)Ben-Nun, Quenneville, and
  Mart{\'\i}nez]{ben2000ab}
Ben-Nun,~M.; Quenneville,~J.; Mart{\'\i}nez,~T.~J. Ab initio multiple spawning:
  Photochemistry from first principles quantum molecular dynamics. \emph{The
  Journal of Physical Chemistry A} \textbf{2000}, \emph{104}, 5161--5175\relax
\mciteBstWouldAddEndPuncttrue
\mciteSetBstMidEndSepPunct{\mcitedefaultmidpunct}
{\mcitedefaultendpunct}{\mcitedefaultseppunct}\relax
\EndOfBibitem
\bibitem[Luehr \latin{et~al.}(2011)Luehr, Ufimtsev, and
  Martinez]{luehr2011dynamic}
Luehr,~N.; Ufimtsev,~I.~S.; Martinez,~T.~J. Dynamic precision for electron
  repulsion integral evaluation on graphical processing units (GPUs).
  \emph{Journal of Chemical Theory and Computation} \textbf{2011}, \emph{7},
  949--954\relax
\mciteBstWouldAddEndPuncttrue
\mciteSetBstMidEndSepPunct{\mcitedefaultmidpunct}
{\mcitedefaultendpunct}{\mcitedefaultseppunct}\relax
\EndOfBibitem
\bibitem[Snyder~Jr \latin{et~al.}(2016)Snyder~Jr, Curchod, and
  Mart{\'\i}nez]{snyder2016gpu}
Snyder~Jr,~J.~W.; Curchod,~B.~F.; Mart{\'\i}nez,~T.~J. GPU-accelerated
  state-averaged complete active space self-consistent field interfaced with ab
  initio multiple spawning unravels the photodynamics of provitamin D3.
  \emph{The journal of physical chemistry letters} \textbf{2016}, \emph{7},
  2444--2449\relax
\mciteBstWouldAddEndPuncttrue
\mciteSetBstMidEndSepPunct{\mcitedefaultmidpunct}
{\mcitedefaultendpunct}{\mcitedefaultseppunct}\relax
\EndOfBibitem
\bibitem[Curchod \latin{et~al.}(2017)Curchod, Sisto, and
  Mart{\'\i}nez]{curchod2017ab}
Curchod,~B.~F.; Sisto,~A.; Mart{\'\i}nez,~T.~J. Ab initio multiple spawning
  photochemical dynamics of DMABN using GPUs. \emph{The Journal of Physical
  Chemistry A} \textbf{2017}, \emph{121}, 265--276\relax
\mciteBstWouldAddEndPuncttrue
\mciteSetBstMidEndSepPunct{\mcitedefaultmidpunct}
{\mcitedefaultendpunct}{\mcitedefaultseppunct}\relax
\EndOfBibitem
\bibitem[Lee \latin{et~al.}(2018)Lee, Cerutti, Mermelstein, Lin, LeGrand,
  Giese, Roitberg, Case, Walker, and York]{lee2018gpu}
Lee,~T.-S.; Cerutti,~D.~S.; Mermelstein,~D.; Lin,~C.; LeGrand,~S.;
  Giese,~T.~J.; Roitberg,~A.; Case,~D.~A.; Walker,~R.~C.; York,~D.~M.
  GPU-accelerated molecular dynamics and free energy methods in Amber18:
  performance enhancements and new features. \emph{Journal of chemical
  information and modeling} \textbf{2018}, \emph{58}, 2043--2050\relax
\mciteBstWouldAddEndPuncttrue
\mciteSetBstMidEndSepPunct{\mcitedefaultmidpunct}
{\mcitedefaultendpunct}{\mcitedefaultseppunct}\relax
\EndOfBibitem
\bibitem[Seritan \latin{et~al.}(2020)Seritan, Bannwarth, Fales, Hohenstein,
  Kokkila-Schumacher, Luehr, Snyder, Song, Titov, Ufimtsev, \latin{et~al.}
  others]{seritan2020terachem}
Seritan,~S.; Bannwarth,~C.; Fales,~B.~S.; Hohenstein,~E.~G.;
  Kokkila-Schumacher,~S.~I.; Luehr,~N.; Snyder,~J.~W.; Song,~C.; Titov,~A.~V.;
  Ufimtsev,~I.~S., \latin{et~al.}  TeraChem: Accelerating electronic structure
  and ab initio molecular dynamics with graphical processing units. \emph{The
  Journal of chemical physics} \textbf{2020}, \emph{152}\relax
\mciteBstWouldAddEndPuncttrue
\mciteSetBstMidEndSepPunct{\mcitedefaultmidpunct}
{\mcitedefaultendpunct}{\mcitedefaultseppunct}\relax
\EndOfBibitem
\bibitem[Kohnke \latin{et~al.}(2020)Kohnke, Kutzner, and
  Grubmüller]{kohnke2020gpu}
Kohnke,~B.; Kutzner,~C.; Grubmüller,~H. A GPU-accelerated fast multipole
  method for GROMACS: performance and accuracy. \emph{Journal of Chemical
  Theory and Computation} \textbf{2020}, \emph{16}, 6938--6949\relax
\mciteBstWouldAddEndPuncttrue
\mciteSetBstMidEndSepPunct{\mcitedefaultmidpunct}
{\mcitedefaultendpunct}{\mcitedefaultseppunct}\relax
\EndOfBibitem
\bibitem[Jones \latin{et~al.}(2022)Jones, Allen, Yang, Drew~Bennett, Gokhale,
  Moshiri, and Rosing]{jones2022accelerators}
Jones,~D.; Allen,~J.~E.; Yang,~Y.; Drew~Bennett,~W.~F.; Gokhale,~M.;
  Moshiri,~N.; Rosing,~T.~S. Accelerators for classical molecular dynamics
  simulations of biomolecules. \emph{Journal of chemical theory and
  computation} \textbf{2022}, \emph{18}, 4047--4069\relax
\mciteBstWouldAddEndPuncttrue
\mciteSetBstMidEndSepPunct{\mcitedefaultmidpunct}
{\mcitedefaultendpunct}{\mcitedefaultseppunct}\relax
\EndOfBibitem
\bibitem[Wang \latin{et~al.}(2024)Wang, Hait, Johnson, Fajen, Zhang, Guerrero,
  and Mart{\'\i}nez]{wang2024extending}
Wang,~Y.; Hait,~D.; Johnson,~K.~G.; Fajen,~O.~J.; Zhang,~J.~H.;
  Guerrero,~R.~D.; Mart{\'\i}nez,~T.~J. Extending GPU-accelerated Gaussian
  integrals in the TeraChem software package to f type orbitals: Implementation
  and applications. \emph{The Journal of Chemical Physics} \textbf{2024},
  \emph{161}\relax
\mciteBstWouldAddEndPuncttrue
\mciteSetBstMidEndSepPunct{\mcitedefaultmidpunct}
{\mcitedefaultendpunct}{\mcitedefaultseppunct}\relax
\EndOfBibitem
\bibitem[Closs \latin{et~al.}(1986)Closs, Calcaterra, Green, Penfield, and
  Miller]{closs:1986:jcp:inverted}
Closs,~G.; Calcaterra,~L.; Green,~N.; Penfield,~K.; Miller,~J. Distance,
  stereoelectronic effects, and the Marcus inverted region in intramolecular
  electron transfer in organic radical anions. \emph{The Journal of Physical
  Chemistry} \textbf{1986}, \emph{90}, 3673--3683\relax
\mciteBstWouldAddEndPuncttrue
\mciteSetBstMidEndSepPunct{\mcitedefaultmidpunct}
{\mcitedefaultendpunct}{\mcitedefaultseppunct}\relax
\EndOfBibitem
\bibitem[Closs \latin{et~al.}(1988)Closs, Piotrowiak, MacInnis, and
  Fleming]{closs1988EET}
Closs,~G.~L.; Piotrowiak,~P.; MacInnis,~J.~M.; Fleming,~G.~R. Determination of
  long-distance intramolecular triplet energy-transfer rates. Quantitative
  comparison with electron transfer. \emph{Journal of the American Chemical
  Society} \textbf{1988}, \emph{110}, 2652--2653\relax
\mciteBstWouldAddEndPuncttrue
\mciteSetBstMidEndSepPunct{\mcitedefaultmidpunct}
{\mcitedefaultendpunct}{\mcitedefaultseppunct}\relax
\EndOfBibitem
\bibitem[Closs \latin{et~al.}(1989)Closs, Johnson, Miller, and
  Piotrowiak]{closs1989dexter}
Closs,~G.~L.; Johnson,~M.~D.; Miller,~J.~R.; Piotrowiak,~P. A connection
  between intramolecular long-range electron, hole, and triplet energy
  transfers. \emph{Journal of the American Chemical Society} \textbf{1989},
  \emph{111}, 3751--3753\relax
\mciteBstWouldAddEndPuncttrue
\mciteSetBstMidEndSepPunct{\mcitedefaultmidpunct}
{\mcitedefaultendpunct}{\mcitedefaultseppunct}\relax
\EndOfBibitem
\bibitem[DeRosa \latin{et~al.}(2025)DeRosa, Qiu, Cofer-Shabica, , and
  Subotnik]{derosa20251dtet}
DeRosa,~J.~R.; Qiu,~T.; Cofer-Shabica,~D.~V.; ; Subotnik,~J.~E. Marcus Theory
  and The Condon Approximation Revisited II: The Horror of Triplet Energy
  Transfer. \textbf{2025}, in preparation\relax
\mciteBstWouldAddEndPuncttrue
\mciteSetBstMidEndSepPunct{\mcitedefaultmidpunct}
{\mcitedefaultendpunct}{\mcitedefaultseppunct}\relax
\EndOfBibitem
\bibitem[Subotnik \latin{et~al.}(2010)Subotnik, Vura-Weis, Sodt, and
  Ratner]{subotnik2010closs}
Subotnik,~J.~E.; Vura-Weis,~J.; Sodt,~A.~J.; Ratner,~M.~A. Predicting accurate
  electronic excitation transfer rates via Marcus theory with Boys or Edmiston-
  Ruedenberg localized diabatization. \emph{The Journal of Physical Chemistry
  A} \textbf{2010}, \emph{114}, 8665--8675\relax
\mciteBstWouldAddEndPuncttrue
\mciteSetBstMidEndSepPunct{\mcitedefaultmidpunct}
{\mcitedefaultendpunct}{\mcitedefaultseppunct}\relax
\EndOfBibitem
\bibitem[Foster and Boys(1960)Foster, and Boys]{foster1960canonical}
Foster,~J.; Boys,~S. Canonical configurational interaction procedure.
  \emph{Reviews of Modern Physics} \textbf{1960}, \emph{32}, 300\relax
\mciteBstWouldAddEndPuncttrue
\mciteSetBstMidEndSepPunct{\mcitedefaultmidpunct}
{\mcitedefaultendpunct}{\mcitedefaultseppunct}\relax
\EndOfBibitem
\bibitem[Subotnik \latin{et~al.}(2008)Subotnik, Yeganeh, Cave, and
  Ratner]{subotnik2008constructing}
Subotnik,~J.~E.; Yeganeh,~S.; Cave,~R.~J.; Ratner,~M.~A. Constructing diabatic
  states from adiabatic states: Extending generalized Mulliken--Hush to
  multiple charge centers with Boys localization. \emph{The Journal of chemical
  physics} \textbf{2008}, \emph{129}\relax
\mciteBstWouldAddEndPuncttrue
\mciteSetBstMidEndSepPunct{\mcitedefaultmidpunct}
{\mcitedefaultendpunct}{\mcitedefaultseppunct}\relax
\EndOfBibitem
\bibitem[Subotnik \latin{et~al.}(2009)Subotnik, Cave, Steele, and
  Shenvi]{subotnik2009initial}
Subotnik,~J.~E.; Cave,~R.~J.; Steele,~R.~P.; Shenvi,~N. The initial and final
  states of electron and energy transfer processes: Diabatization as motivated
  by system-solvent interactions. \emph{The Journal of chemical physics}
  \textbf{2009}, \emph{130}\relax
\mciteBstWouldAddEndPuncttrue
\mciteSetBstMidEndSepPunct{\mcitedefaultmidpunct}
{\mcitedefaultendpunct}{\mcitedefaultseppunct}\relax
\EndOfBibitem
\bibitem[Landry and Subotnik(2014)Landry, and
  Subotnik]{landry:2014:abinitio_closs}
Landry,~B.~R.; Subotnik,~J.~E. Quantifying the lifetime of triplet energy
  transfer processes in organic chromophores: A case study of
  4-(2-naphthylmethyl)benzaldehyde. \emph{Journal of Chemical Theory and
  Computation} \textbf{2014}, \emph{10}, 4253--4263\relax
\mciteBstWouldAddEndPuncttrue
\mciteSetBstMidEndSepPunct{\mcitedefaultmidpunct}
{\mcitedefaultendpunct}{\mcitedefaultseppunct}\relax
\EndOfBibitem
\bibitem[Pacher \latin{et~al.}(1988)Pacher, Cederbaum, and
  K{\"o}ppel]{pacher1988approximately}
Pacher,~T.; Cederbaum,~L.; K{\"o}ppel,~H. Approximately diabatic states from
  block diagonalization of the electronic Hamiltonian. \emph{The Journal of
  chemical physics} \textbf{1988}, \emph{89}, 7367--7381\relax
\mciteBstWouldAddEndPuncttrue
\mciteSetBstMidEndSepPunct{\mcitedefaultmidpunct}
{\mcitedefaultendpunct}{\mcitedefaultseppunct}\relax
\EndOfBibitem
\bibitem[Pacher \latin{et~al.}(1993)Pacher, Cederbaum, and
  K{\"o}ppel]{pacher1993adiabatic}
Pacher,~T.; Cederbaum,~L.; K{\"o}ppel,~H. Adiabatic and quasidiabatic states in
  a gauge theoretical framework. \emph{Advances in chemical physics}
  \textbf{1993}, \emph{84}, 293--391\relax
\mciteBstWouldAddEndPuncttrue
\mciteSetBstMidEndSepPunct{\mcitedefaultmidpunct}
{\mcitedefaultendpunct}{\mcitedefaultseppunct}\relax
\EndOfBibitem
\bibitem[Mead and Truhlar(1982)Mead, and Truhlar]{mead1982conditions}
Mead,~C.~A.; Truhlar,~D.~G. Conditions for the definition of a strictly
  diabatic electronic basis for molecular systems. \emph{The Journal of
  Chemical Physics} \textbf{1982}, \emph{77}, 6090--6098\relax
\mciteBstWouldAddEndPuncttrue
\mciteSetBstMidEndSepPunct{\mcitedefaultmidpunct}
{\mcitedefaultendpunct}{\mcitedefaultseppunct}\relax
\EndOfBibitem
\bibitem[Subotnik \latin{et~al.}(2008)Subotnik, Yeganeh, Cave, and
  Ratner]{subotnik2008boys}
Subotnik,~J.~E.; Yeganeh,~S.; Cave,~R.~J.; Ratner,~M.~A. Constructing diabatic
  states from adiabatic states: Extending generalized Mulliken--Hush to
  multiple charge centers with Boys localization. \emph{The Journal of chemical
  physics} \textbf{2008}, \emph{129}\relax
\mciteBstWouldAddEndPuncttrue
\mciteSetBstMidEndSepPunct{\mcitedefaultmidpunct}
{\mcitedefaultendpunct}{\mcitedefaultseppunct}\relax
\EndOfBibitem
\bibitem[Cave and Newton(1996)Cave, and Newton]{cave1996generalization}
Cave,~R.~J.; Newton,~M.~D. Generalization of the Mulliken-Hush treatment for
  the calculation of electron transfer matrix elements. \emph{Chemical physics
  letters} \textbf{1996}, \emph{249}, 15--19\relax
\mciteBstWouldAddEndPuncttrue
\mciteSetBstMidEndSepPunct{\mcitedefaultmidpunct}
{\mcitedefaultendpunct}{\mcitedefaultseppunct}\relax
\EndOfBibitem
\bibitem[Voityuk and R{\"o}sch(2002)Voityuk, and
  R{\"o}sch]{voityuk2002fragment}
Voityuk,~A.~A.; R{\"o}sch,~N. Fragment charge difference method for estimating
  donor--acceptor electronic coupling: Application to DNA $\pi$-stacks.
  \emph{The Journal of chemical physics} \textbf{2002}, \emph{117},
  5607--5616\relax
\mciteBstWouldAddEndPuncttrue
\mciteSetBstMidEndSepPunct{\mcitedefaultmidpunct}
{\mcitedefaultendpunct}{\mcitedefaultseppunct}\relax
\EndOfBibitem
\bibitem[Hsu \latin{et~al.}(2008)Hsu, You, and Chen]{hsu2008characterization}
Hsu,~C.-P.; You,~Z.-Q.; Chen,~H.-C. Characterization of the short-range
  couplings in excitation energy transfer. \emph{The Journal of Physical
  Chemistry C} \textbf{2008}, \emph{112}, 1204--1212\relax
\mciteBstWouldAddEndPuncttrue
\mciteSetBstMidEndSepPunct{\mcitedefaultmidpunct}
{\mcitedefaultendpunct}{\mcitedefaultseppunct}\relax
\EndOfBibitem
\bibitem[Paz and Glover(2023)Paz, and Glover]{glover2023diabatgradients}
Paz,~A.~S.; Glover,~W.~J. Efficient analytical gradients of property-based
  diabatic states: Geometry optimizations for localized holes. \emph{The
  Journal of Chemical Physics} \textbf{2023}, \emph{158}\relax
\mciteBstWouldAddEndPuncttrue
\mciteSetBstMidEndSepPunct{\mcitedefaultmidpunct}
{\mcitedefaultendpunct}{\mcitedefaultseppunct}\relax
\EndOfBibitem
\bibitem[Athavale \latin{et~al.}(2025)Athavale, Subotnik, and
  Qiu]{athavale2025evaluating}
Athavale,~V.; Subotnik,~J.~E.; Qiu,~T. Evaluating the gradients of localized
  diabatic state energies and couplings at minimum cost. \emph{The Journal of
  Chemical Physics} \textbf{2025}, \emph{163}\relax
\mciteBstWouldAddEndPuncttrue
\mciteSetBstMidEndSepPunct{\mcitedefaultmidpunct}
{\mcitedefaultendpunct}{\mcitedefaultseppunct}\relax
\EndOfBibitem
\bibitem[Fatehi \latin{et~al.}(2013)Fatehi, Alguire, and
  Subotnik]{fatehi2013diabatderivatives}
Fatehi,~S.; Alguire,~E.; Subotnik,~J.~E. Derivative couplings and analytic
  gradients for diabatic states, with an implementation for Boys-localized
  configuration-interaction singles. \emph{The Journal of Chemical Physics}
  \textbf{2013}, \emph{139}\relax
\mciteBstWouldAddEndPuncttrue
\mciteSetBstMidEndSepPunct{\mcitedefaultmidpunct}
{\mcitedefaultendpunct}{\mcitedefaultseppunct}\relax
\EndOfBibitem
\bibitem[Coffman \latin{et~al.}(2023)Coffman, Jin, Chen, Subotnik, and
  Cofer-Shabica]{memes}
Coffman,~A.~J.; Jin,~Z.; Chen,~J.; Subotnik,~J.~E.; Cofer-Shabica,~D.~V. Use of
  QM/MM Surface Hopping Simulations to Understand Thermally Activated
  Rare-Event Nonadiabatic Transitions in the Condensed Phase. \emph{Journal of
  Chemical Theory and Computation} \textbf{2023}, \emph{19}, 7136--7150\relax
\mciteBstWouldAddEndPuncttrue
\mciteSetBstMidEndSepPunct{\mcitedefaultmidpunct}
{\mcitedefaultendpunct}{\mcitedefaultseppunct}\relax
\EndOfBibitem
\bibitem[Lindner \latin{et~al.}(2019)Lindner, Sultangaleeva, R{\"o}hr, and
  Mitri{\'c}]{metafalcon}
Lindner,~J.~O.; Sultangaleeva,~K.; R{\"o}hr,~M. I.~S.; Mitri{\'c},~R.
  {{metaFALCON}}: {{A Program Package}} for {{Automatic Sampling}} of {{Conical
  Intersection Seams Using Multistate Metadynamics}}. \emph{Journal of Chemical
  Theory and Computation} \textbf{2019}, \emph{15}, 3450--3460\relax
\mciteBstWouldAddEndPuncttrue
\mciteSetBstMidEndSepPunct{\mcitedefaultmidpunct}
{\mcitedefaultendpunct}{\mcitedefaultseppunct}\relax
\EndOfBibitem
\bibitem[Mori and Mart{\'i}nez(2013)Mori, and Mart{\'i}nez]{mori2013nebci}
Mori,~T.; Mart{\'i}nez,~{\relax Todd}.~J. Exploring the {{Conical Intersection
  Seam}}: {{The Seam Space Nudged Elastic Band Method}}. \emph{Journal of
  Chemical Theory and Computation} \textbf{2013}, \emph{9}, 1155--1163\relax
\mciteBstWouldAddEndPuncttrue
\mciteSetBstMidEndSepPunct{\mcitedefaultmidpunct}
{\mcitedefaultendpunct}{\mcitedefaultseppunct}\relax
\EndOfBibitem
\bibitem[Andersen(1983)]{andersen1983RattleVelocityVersion}
Andersen,~H.~C. Rattle: {A} “velocity” version of the shake algorithm for
  molecular dynamics calculations. \emph{Journal of Computational Physics}
  \textbf{1983}, \emph{52}, 24--34\relax
\mciteBstWouldAddEndPuncttrue
\mciteSetBstMidEndSepPunct{\mcitedefaultmidpunct}
{\mcitedefaultendpunct}{\mcitedefaultseppunct}\relax
\EndOfBibitem
\bibitem[Cofer-Shabica \latin{et~al.}(2022)Cofer-Shabica, Menger, Ou, Shao,
  Subotnik, and Faraji]{inaqs:1}
Cofer-Shabica,~D.~V.; Menger,~M. F. S.~J.; Ou,~Q.; Shao,~Y.; Subotnik,~J.~E.;
  Faraji,~S. INAQS, a Generic Interface for Nonadiabatic QM/MM Dynamics:
  Design, Implementation, and Validation for GROMACS/Q-CHEM simulations.
  \emph{Journal of Chemical Theory and Computation} \textbf{2022}, \emph{18},
  4601--4614\relax
\mciteBstWouldAddEndPuncttrue
\mciteSetBstMidEndSepPunct{\mcitedefaultmidpunct}
{\mcitedefaultendpunct}{\mcitedefaultseppunct}\relax
\EndOfBibitem
\bibitem[Pronk \latin{et~al.}(2013)Pronk, P\'{a}ll, Schulz, Larsson, Bjelkmar,
  Apostolov, Shirts, Smith, Kasson, van~der Spoel, Hess, and
  Lindahl]{gromacs45}
Pronk,~S.; P\'{a}ll,~S.; Schulz,~R.; Larsson,~P.; Bjelkmar,~P.; Apostolov,~R.;
  Shirts,~M.~R.; Smith,~J.~C.; Kasson,~P.~M.; van~der Spoel,~D.; Hess,~B.;
  Lindahl,~E. {GROMACS 4.5: a high-throughput and highly parallel open source
  molecular simulation toolkit}. \emph{Bioinformatics} \textbf{2013},
  \emph{29}, 845--854\relax
\mciteBstWouldAddEndPuncttrue
\mciteSetBstMidEndSepPunct{\mcitedefaultmidpunct}
{\mcitedefaultendpunct}{\mcitedefaultseppunct}\relax
\EndOfBibitem
\bibitem[DeRosa \latin{et~al.}(2025)DeRosa, Subotnik, Pei, Shao, Shuman, Ard,
  Viggiano, and Cofer-Shabica]{derosa2024seam}
DeRosa,~J.~R.; Subotnik,~J.~E.; Pei,~Z.; Shao,~Y.; Shuman,~N.~S.; Ard,~S.~G.;
  Viggiano,~A.~A.; Cofer-Shabica,~D.~V. Revisiting the Discrepancy Between
  Experimental and Theoretical Predictions of the Adiabaticity of {Ti$^+$ +
  CH$_3$OH}. \emph{The Journal of Physical Chemistry A} \textbf{2025}, \relax
\mciteBstWouldAddEndPunctfalse
\mciteSetBstMidEndSepPunct{\mcitedefaultmidpunct}
{}{\mcitedefaultseppunct}\relax
\EndOfBibitem
\bibitem[Verlet(1967)]{verlet}
Verlet,~L. Computer" experiments" on classical fluids. I. Thermodynamical
  properties of Lennard-Jones molecules. \emph{Physical review} \textbf{1967},
  \emph{159}, 98\relax
\mciteBstWouldAddEndPuncttrue
\mciteSetBstMidEndSepPunct{\mcitedefaultmidpunct}
{\mcitedefaultendpunct}{\mcitedefaultseppunct}\relax
\EndOfBibitem
\bibitem[Swope \latin{et~al.}(1982)Swope, Andersen, Berens, and
  Wilson]{vverlet}
Swope,~W.~C.; Andersen,~H.~C.; Berens,~P.~H.; Wilson,~K.~R. A computer
  simulation method for the calculation of equilibrium constants for the
  formation of physical clusters of molecules: Application to small water
  clusters. \emph{The Journal of chemical physics} \textbf{1982}, \emph{76},
  637--649\relax
\mciteBstWouldAddEndPuncttrue
\mciteSetBstMidEndSepPunct{\mcitedefaultmidpunct}
{\mcitedefaultendpunct}{\mcitedefaultseppunct}\relax
\EndOfBibitem
\bibitem[val()]{vale_footnote}
Extension to sampling conical intersections---seams between
  \emph{adiabats}---is easy, though the solution to \refeq{eq:const:g} must be
  found via a 1-dimensional minimization rather than a root search because the
  energy difference between 2 successive adiabats is never negative.\relax
\mciteBstWouldAddEndPunctfalse
\mciteSetBstMidEndSepPunct{\mcitedefaultmidpunct}
{}{\mcitedefaultseppunct}\relax
\EndOfBibitem
\bibitem[Press(2007)]{press2007}
Press,~W.~H., Ed. \emph{Numerical Recipes: The Art of Scientific Computing},
  3rd ed.; Cambridge University Press: Cambridge, UK ; New York, 2007\relax
\mciteBstWouldAddEndPuncttrue
\mciteSetBstMidEndSepPunct{\mcitedefaultmidpunct}
{\mcitedefaultendpunct}{\mcitedefaultseppunct}\relax
\EndOfBibitem
\bibitem[Epifanovsky \latin{et~al.}(2021)Epifanovsky, Gilbert, Feng, Lee, Mao,
  Mardirossian, Pokhilko, White, Coons, Dempwolff, Gan, Hait, Horn, Jacobson,
  Kaliman, Kussmann, Lange, Lao, Levine, Liu, McKenzie, Morrison, Nanda,
  Plasser, Rehn, Vidal, You, Zhu, Alam, Albrecht, Aldossary, Alguire, Andersen,
  Athavale, Barton, Begam, Behn, Bellonzi, Bernard, Berquist, Burton, Carreras,
  Carter-Fenk, Chakraborty, Chien, Closser, Cofer-Shabica, Dasgupta,
  de~Wergifosse, Deng, Diedenhofen, Do, Ehlert, Fang, Fatehi, Feng, Friedhoff,
  Gayvert, Ge, Gidofalvi, Goldey, Gomes, Gonz\'{a}lez-Espinoza, Gulania,
  Gunina, Hanson-Heine, Harbach, Hauser, Herbst, Hernández~Vera, Hodecker,
  Holden, Houck, Huang, Hui, Huynh, Ivanov, J\'{a}sz, Ji, Jiang, Kaduk,
  K\"{a}hler, Khistyaev, Kim, Kis, Klunzinger, Koczor-Benda, Koh, Kosenkov,
  Koulias, Kowalczyk, Krauter, Kue, Kunitsa, Kus, Ladj\'{a}nszki, Landau,
  Lawler, Lefrancois, Lehtola, Li, Li, Liang, Liebenthal, Lin, Lin, Liu, Liu,
  Loipersberger, Luenser, Manjanath, Manohar, Mansoor, Manzer, Mao, Marenich,
  Markovich, Mason, Maurer, McLaughlin, Menger, Mewes, Mewes, Morgante,
  Mullinax, Oosterbaan, Paran, Paul, Paul, Pavo\v{s}evi\'{c}, Pei, Prager,
  Proynov, R\'{a}k, Ramos-Cordoba, Rana, Rask, Rettig, Richard, Rob, Rossomme,
  Scheele, Scheurer, Schneider, Sergueev, Sharada, Skomorowski, Small, Stein,
  Su, Sundstrom, Tao, Thirman, Tornai, Tsuchimochi, Tubman, Veccham, Vydrov,
  Wenzel, Witte, Yamada, Yao, Yeganeh, Yost, Zech, Zhang, Zhang, Zhang, Zuev,
  Aspuru-Guzik, Bell, Besley, Bravaya, Brooks, Casanova, Chai, Coriani, Cramer,
  Cserey, DePrince, DiStasio, Dreuw, Dunietz, Furlani, Goddard,
  Hammes-Schiffer, Head-Gordon, Hehre, Hsu, Jagau, Jung, Klamt, Kong,
  Lambrecht, Liang, Mayhall, McCurdy, Neaton, Ochsenfeld, Parkhill, Peverati,
  Rassolov, Shao, Slipchenko, Stauch, Steele, Subotnik, Thom, Tkatchenko,
  Truhlar, Van~Voorhis, Wesolowski, Whaley, Woodcock, Zimmerman, Faraji, Gill,
  Head-Gordon, Herbert, and Krylov]{qchem}
Epifanovsky,~E.; Gilbert,~A. T.~B.; Feng,~X.; Lee,~J.; Mao,~Y.;
  Mardirossian,~N.; Pokhilko,~P.; White,~A.~F.; Coons,~M.~P.; Dempwolff,~A.~L.;
  Gan,~Z.; Hait,~D.; Horn,~P.~R.; Jacobson,~L.~D.; Kaliman,~I.; Kussmann,~J.;
  Lange,~A.~W.; Lao,~K.~U.; Levine,~D.~S.; Liu,~J.; McKenzie,~S.~C.;
  Morrison,~A.~F.; Nanda,~K.~D.; Plasser,~F.; Rehn,~D.~R.; Vidal,~M.~L.;
  You,~Z.-Q.; Zhu,~Y.; Alam,~B.; Albrecht,~B.~J.; Aldossary,~A.; Alguire,~E.;
  Andersen,~J.~H.; Athavale,~V.; Barton,~D.; Begam,~K.; Behn,~A.; Bellonzi,~N.;
  Bernard,~Y.~A.; Berquist,~E.~J.; Burton,~H. G.~A.; Carreras,~A.;
  Carter-Fenk,~K.; Chakraborty,~R.; Chien,~A.~D.; Closser,~K.~D.;
  Cofer-Shabica,~V.; Dasgupta,~S.; de~Wergifosse,~M.; Deng,~J.;
  Diedenhofen,~M.; Do,~H.; Ehlert,~S.; Fang,~P.-T.; Fatehi,~S.; Feng,~Q.;
  Friedhoff,~T.; Gayvert,~J.; Ge,~Q.; Gidofalvi,~G.; Goldey,~M.; Gomes,~J.;
  Gonz\'{a}lez-Espinoza,~C.~E.; Gulania,~S.; Gunina,~A.~O.; Hanson-Heine,~M.
  W.~D.; Harbach,~P. H.~P.; Hauser,~A.; Herbst,~M.~F.; Hernández~Vera,~M.;
  Hodecker,~M.; Holden,~Z.~C.; Houck,~S.; Huang,~X.; Hui,~K.; Huynh,~B.~C.;
  Ivanov,~M.; J\'{a}sz,~A.; Ji,~H.; Jiang,~H.; Kaduk,~B.; K\"{a}hler,~S.;
  Khistyaev,~K.; Kim,~J.; Kis,~G.; Klunzinger,~P.; Koczor-Benda,~Z.;
  Koh,~J.~H.; Kosenkov,~D.; Koulias,~L.; Kowalczyk,~T.; Krauter,~C.~M.;
  Kue,~K.; Kunitsa,~A.; Kus,~T.; Ladj\'{a}nszki,~I.; Landau,~A.; Lawler,~K.~V.;
  Lefrancois,~D.; Lehtola,~S.; Li,~R.~R.; Li,~Y.-P.; Liang,~J.; Liebenthal,~M.;
  Lin,~H.-H.; Lin,~Y.-S.; Liu,~F.; Liu,~K.-Y.; Loipersberger,~M.; Luenser,~A.;
  Manjanath,~A.; Manohar,~P.; Mansoor,~E.; Manzer,~S.~F.; Mao,~S.-P.;
  Marenich,~A.~V.; Markovich,~T.; Mason,~S.; Maurer,~S.~A.; McLaughlin,~P.~F.;
  Menger,~M. F. S.~J.; Mewes,~J.-M.; Mewes,~S.~A.; Morgante,~P.;
  Mullinax,~J.~W.; Oosterbaan,~K.~J.; Paran,~G.; Paul,~A.~C.; Paul,~S.~K.;
  Pavo\v{s}evi\'{c},~F.; Pei,~Z.; Prager,~S.; Proynov,~E.~I.; R\'{a}k,~A.;
  Ramos-Cordoba,~E.; Rana,~B.; Rask,~A.~E.; Rettig,~A.; Richard,~R.~M.;
  Rob,~F.; Rossomme,~E.; Scheele,~T.; Scheurer,~M.; Schneider,~M.;
  Sergueev,~N.; Sharada,~S.~M.; Skomorowski,~W.; Small,~D.~W.; Stein,~C.~J.;
  Su,~Y.-C.; Sundstrom,~E.~J.; Tao,~Z.; Thirman,~J.; Tornai,~G.~J.;
  Tsuchimochi,~T.; Tubman,~N.~M.; Veccham,~S.~P.; Vydrov,~O.; Wenzel,~J.;
  Witte,~J.; Yamada,~A.; Yao,~K.; Yeganeh,~S.; Yost,~S.~R.; Zech,~A.;
  Zhang,~I.~Y.; Zhang,~X.; Zhang,~Y.; Zuev,~D.; Aspuru-Guzik,~A.; Bell,~A.~T.;
  Besley,~N.~A.; Bravaya,~K.~B.; Brooks,~B.~R.; Casanova,~D.; Chai,~J.-D.;
  Coriani,~S.; Cramer,~C.~J.; Cserey,~G.; DePrince,~A.~E.; DiStasio,~R.~A.;
  Dreuw,~A.; Dunietz,~B.~D.; Furlani,~T.~R.; Goddard,~W.~A.;
  Hammes-Schiffer,~S.; Head-Gordon,~T.; Hehre,~W.~J.; Hsu,~C.-P.; Jagau,~T.-C.;
  Jung,~Y.; Klamt,~A.; Kong,~J.; Lambrecht,~D.~S.; Liang,~W.; Mayhall,~N.~J.;
  McCurdy,~C.~W.; Neaton,~J.~B.; Ochsenfeld,~C.; Parkhill,~J.~A.; Peverati,~R.;
  Rassolov,~V.~A.; Shao,~Y.; Slipchenko,~L.~V.; Stauch,~T.; Steele,~R.~P.;
  Subotnik,~J.~E.; Thom,~A. J.~W.; Tkatchenko,~A.; Truhlar,~D.~G.;
  Van~Voorhis,~T.; Wesolowski,~T.~A.; Whaley,~K.~B.; Woodcock,~H.~L.;
  Zimmerman,~P.~M.; Faraji,~S.; Gill,~P. M.~W.; Head-Gordon,~M.;
  Herbert,~J.~M.; Krylov,~A.~I. Software for the frontiers of quantum
  chemistry: An overview of developments in the Q-Chem 5 package. \emph{The
  Journal of Chemical Physics} \textbf{2021}, \emph{155}, 084801\relax
\mciteBstWouldAddEndPuncttrue
\mciteSetBstMidEndSepPunct{\mcitedefaultmidpunct}
{\mcitedefaultendpunct}{\mcitedefaultseppunct}\relax
\EndOfBibitem
\bibitem[Bussi \latin{et~al.}(2007)Bussi, Donadio, and
  Parrinello]{bussi2007thermostat}
Bussi,~G.; Donadio,~D.; Parrinello,~M. Canonical sampling through velocity
  rescaling. \emph{The Journal of chemical physics} \textbf{2007},
  \emph{126}\relax
\mciteBstWouldAddEndPuncttrue
\mciteSetBstMidEndSepPunct{\mcitedefaultmidpunct}
{\mcitedefaultendpunct}{\mcitedefaultseppunct}\relax
\EndOfBibitem
\bibitem[Moser \latin{et~al.}(1992)Moser, Keske, Warncke, Farid, and
  Dutton]{moser1992nature}
Moser,~C.~C.; Keske,~J.~M.; Warncke,~K.; Farid,~R.~S.; Dutton,~P.~L. Nature of
  biological electron transfer. \emph{Nature} \textbf{1992}, \emph{355},
  796--802\relax
\mciteBstWouldAddEndPuncttrue
\mciteSetBstMidEndSepPunct{\mcitedefaultmidpunct}
{\mcitedefaultendpunct}{\mcitedefaultseppunct}\relax
\EndOfBibitem
\bibitem[Behn \latin{et~al.}(2011)Behn, Zimmerman, Bell, and
  Head-Gordon]{behn2011fsm}
Behn,~A.; Zimmerman,~P.~M.; Bell,~A.~T.; Head-Gordon,~M. Efficient exploration
  of reaction paths via a freezing string method. \emph{The Journal of chemical
  physics} \textbf{2011}, \emph{135}\relax
\mciteBstWouldAddEndPuncttrue
\mciteSetBstMidEndSepPunct{\mcitedefaultmidpunct}
{\mcitedefaultendpunct}{\mcitedefaultseppunct}\relax
\EndOfBibitem
\bibitem[Head-Gordon \latin{et~al.}(1995)Head-Gordon, Grana, Maurice, and
  White]{mhg1995attachdetach}
Head-Gordon,~M.; Grana,~A.~M.; Maurice,~D.; White,~C.~A. Analysis of electronic
  transitions as the difference of electron attachment and detachment
  densities. \emph{The Journal of Physical Chemistry} \textbf{1995}, \emph{99},
  14261--14270\relax
\mciteBstWouldAddEndPuncttrue
\mciteSetBstMidEndSepPunct{\mcitedefaultmidpunct}
{\mcitedefaultendpunct}{\mcitedefaultseppunct}\relax
\EndOfBibitem
\bibitem[Yarkony(1996)]{yarkony1996diabolical}
Yarkony,~D.~R. Diabolical conical intersections. \emph{Reviews of Modern
  Physics} \textbf{1996}, \emph{68}, 985\relax
\mciteBstWouldAddEndPuncttrue
\mciteSetBstMidEndSepPunct{\mcitedefaultmidpunct}
{\mcitedefaultendpunct}{\mcitedefaultseppunct}\relax
\EndOfBibitem
\bibitem[Manaa and Yarkony(1993)Manaa, and Yarkony]{manaa1993intersection}
Manaa,~M.~R.; Yarkony,~D.~R. On the intersection of two potential energy
  surfaces of the same symmetry. Systematic characterization using a Lagrange
  multiplier constrained procedure. \emph{The Journal of chemical physics}
  \textbf{1993}, \emph{99}, 5251--5256\relax
\mciteBstWouldAddEndPuncttrue
\mciteSetBstMidEndSepPunct{\mcitedefaultmidpunct}
{\mcitedefaultendpunct}{\mcitedefaultseppunct}\relax
\EndOfBibitem
\bibitem[Bearpark \latin{et~al.}(1994)Bearpark, Robb, and
  Schlegel]{bearpark1994direct}
Bearpark,~M.~J.; Robb,~M.~A.; Schlegel,~H.~B. A direct method for the location
  of the lowest energy point on a potential surface crossing. \emph{Chemical
  physics letters} \textbf{1994}, \emph{223}, 269--274\relax
\mciteBstWouldAddEndPuncttrue
\mciteSetBstMidEndSepPunct{\mcitedefaultmidpunct}
{\mcitedefaultendpunct}{\mcitedefaultseppunct}\relax
\EndOfBibitem
\bibitem[Domcke and Yarkony(2012)Domcke, and Yarkony]{domcke2012role}
Domcke,~W.; Yarkony,~D.~R. Role of conical intersections in molecular
  spectroscopy and photoinduced chemical dynamics. \emph{Annual review of
  physical chemistry} \textbf{2012}, \emph{63}, 325--352\relax
\mciteBstWouldAddEndPuncttrue
\mciteSetBstMidEndSepPunct{\mcitedefaultmidpunct}
{\mcitedefaultendpunct}{\mcitedefaultseppunct}\relax
\EndOfBibitem
\bibitem[Fu{\ss} \latin{et~al.}(2002)Fu{\ss}, Pushpa, Rettig, Schmid, and
  Trushin]{fuss2002ultrafast}
Fu{\ss},~W.; Pushpa,~K.~K.; Rettig,~W.; Schmid,~W.~E.; Trushin,~S.~A. Ultrafast
  charge transfer via a conical intersection in dimethylaminobenzonitrile.
  \emph{Photochemical \& Photobiological Sciences} \textbf{2002}, \emph{1},
  255--262\relax
\mciteBstWouldAddEndPuncttrue
\mciteSetBstMidEndSepPunct{\mcitedefaultmidpunct}
{\mcitedefaultendpunct}{\mcitedefaultseppunct}\relax
\EndOfBibitem
\bibitem[Rettig(1986)]{rettig1986charge}
Rettig,~W. Charge separation in excited states of decoupled systems—TICT
  compounds and implications regarding the development of new laser dyes and
  the primary process of vision and photosynthesis. \emph{Angewandte Chemie
  International Edition in English} \textbf{1986}, \emph{25}, 971--988\relax
\mciteBstWouldAddEndPuncttrue
\mciteSetBstMidEndSepPunct{\mcitedefaultmidpunct}
{\mcitedefaultendpunct}{\mcitedefaultseppunct}\relax
\EndOfBibitem
\bibitem[Rettig(1992)]{rettig1992photoinduced}
Rettig,~W. Photoinduced electron transfer in twisted $\pi$-systems.
  \emph{Dynamics and mechanisms of Photoinduced Electron Transfer and Related
  Phenomena} \textbf{1992}, 57--69\relax
\mciteBstWouldAddEndPuncttrue
\mciteSetBstMidEndSepPunct{\mcitedefaultmidpunct}
{\mcitedefaultendpunct}{\mcitedefaultseppunct}\relax
\EndOfBibitem
\bibitem[Hsu(2009)]{hsu2009electronic}
Hsu,~C.-P. The electronic couplings in electron transfer and excitation energy
  transfer. \emph{Accounts of Chemical Research} \textbf{2009}, \emph{42},
  509--518\relax
\mciteBstWouldAddEndPuncttrue
\mciteSetBstMidEndSepPunct{\mcitedefaultmidpunct}
{\mcitedefaultendpunct}{\mcitedefaultseppunct}\relax
\EndOfBibitem
\bibitem[Krueger \latin{et~al.}(1998)Krueger, Scholes, and
  Fleming]{krueger1998calculation}
Krueger,~B.~P.; Scholes,~G.~D.; Fleming,~G.~R. Calculation of couplings and
  energy-transfer pathways between the pigments of LH2 by the ab initio
  transition density cube method. \emph{The Journal of Physical Chemistry B}
  \textbf{1998}, \emph{102}, 5378--5386\relax
\mciteBstWouldAddEndPuncttrue
\mciteSetBstMidEndSepPunct{\mcitedefaultmidpunct}
{\mcitedefaultendpunct}{\mcitedefaultseppunct}\relax
\EndOfBibitem
\bibitem[Hsu \latin{et~al.}(2001)Hsu, Walla, Head-Gordon, and
  Fleming]{hsu2001role}
Hsu,~C.-P.; Walla,~P.~J.; Head-Gordon,~M.; Fleming,~G.~R. The role of the S1
  state of carotenoids in photosynthetic energy transfer: the light-harvesting
  complex II of purple bacteria. \emph{The Journal of Physical Chemistry B}
  \textbf{2001}, \emph{105}, 11016--11025\relax
\mciteBstWouldAddEndPuncttrue
\mciteSetBstMidEndSepPunct{\mcitedefaultmidpunct}
{\mcitedefaultendpunct}{\mcitedefaultseppunct}\relax
\EndOfBibitem
\bibitem[Krausz and Ivanov(2009)Krausz, and Ivanov]{krausz2009attosecond}
Krausz,~F.; Ivanov,~M. Attosecond physics. \emph{Reviews of modern physics}
  \textbf{2009}, \emph{81}, 163--234\relax
\mciteBstWouldAddEndPuncttrue
\mciteSetBstMidEndSepPunct{\mcitedefaultmidpunct}
{\mcitedefaultendpunct}{\mcitedefaultseppunct}\relax
\EndOfBibitem
\end{mcitethebibliography}

\end{document}